\documentclass[fleqn,usenatbib]{mnras}

\usepackage{newtxtext,newtxmath}
 
\usepackage[T1]{fontenc}

\DeclareRobustCommand{\VAN}[3]{#2}
\let\VANthebibliography\thebibliography
\def\thebibliography{\DeclareRobustCommand{\VAN}[3]{##3}\VANthebibliography}

\usepackage{graphicx}	
\usepackage{amsmath}	
\usepackage{amssymb}	

\usepackage[capitalise]{cleveref} 

\usepackage[caption=false]{subfig}
\usepackage{listings} 

\newcommand{\pccm}{{\rm pc~cm}^{-3}}

\newcommand{\HI}{\ion{H}{i}}
\newcommand{\HII}{\ion{H}{ii}}
\newcommand{\HeI}{\ion{He}{i}}
\newcommand{\HeII}{\ion{He}{ii}}
\newcommand{\HeIII}{\ion{He}{iii}}

\newcommand{\EAGLE}{\mbox{EAGLE}}

\newcommand{\DMcosmic}{$\mathrm{DM_{cosmic}}$}

\crefname{lstlisting}{listing}{listings}
\Crefname{lstlisting}{Listing}{Listings}

\title[The Cosmic Dispersion Measure in EAGLE]{The Cosmic Dispersion Measure in the EAGLE Simulations}

\author[A. J. Batten et al.]{
Adam J. Batten,$^{1,2}$\thanks{E-mail: abatten@swin.edu.au}
Alan R. Duffy,$^{1,2}$ 
Nastasha A. Wijers,$^{3}$
Vivek Gupta,$^{1}$
Chris Flynn,$^{1}$
\newauthor 
Joop Schaye,$^{3}$ 
Emma Ryan-Weber$^{1,2}$
\\
\\
$^{1}$Centre for Astrophysics and Supercomputing, Swinburne University of Technology, Melbourne, Victoria 3122, Australia \\
$^{2}$ARC Centre of Excellence for All Sky Astrophysics in 3 Dimensions (ASTRO 3D) \\
$^{3}$Sterrewacht, Leiden University, Niels Bohrweg 2, 2333 CA Leiden, The Netherlands\\
}

\date{Accepted XXX. Received YYY; in original form ZZZ}

\pubyear{2020}

\begin{document}
\label{firstpage}
\pagerange{\pageref{firstpage}--\pageref{lastpage}}
\maketitle

\begin{abstract}
The dispersion measure (DM) of fast radio bursts (FRBs) provides a unique way to probe ionised baryons in the intergalactic medium (IGM).
Cosmological models with different parameters lead to different DM-redshift ($\mathrm{DM}-z$) relations.
Additionally, the over/under-dense regions in the IGM and the circumgalactic medium of intervening galaxies lead to scatter around the mean $\mathrm{DM}-z$ relations. 
We have used the Evolution and Assembly of GaLaxies and their Environments (\EAGLE) simulations to measure the mean $\mathrm{DM}-z$ relation and the scatter around it using over one billion lines-of-sight between redshifts $0<z<3$. 
We investigated two techniques to estimate line-of-sight DM: `pixel scrambling' and `box transformations'. 
We find that using box transformations (a technique from the literature) causes strong correlations due to repeated replication of structure.
Comparing a linear and non-linear model, we find that the non-linear model with cosmological parameters, provides a better fit to the $\mathrm{DM}-z$ relation. The differences between these models are the most significant at low redshifts ($z<0.5$).
The scatter around the $\mathrm{DM}-z$ relation is highly asymmetric, especially at low redshift $\left(z<0.5\right)$, and becomes more Gaussian as redshift approaches $z\sim3$, the limit of this study.
The increase in Gaussianity with redshift is indicative of the large scale structures that is better probed with longer lines-of-sight.
The minimum simulation size suitable for investigations into the scatter around the $\mathrm{DM}-z$ relation is 100~comoving~Mpc.
The $\mathrm{DM}-z$ relation measured in \EAGLE\ is available with an easy-to-use python interface in the open-source FRB redshift estimation package \textsc{fruitbat}.

\end{abstract}


\begin{keywords}
intergalactic medium -- hydrodynamics -- methods: numerical -- radio continuum: general 
\end{keywords}



\section{Introduction}
The whereabouts of almost a third of the baryons in the Universe is still unknown. A census by \citet{Shull2012} to identify the baryonic content of the Universe at redshift $z=0$ found that $29 \pm 13\%$ remains undetected. This is known as the `missing baryon problem'.


Cosmological hydrodynamic simulations of a lambda cold dark matter ($\Lambda$CDM) Universe predict that these `missing baryons' reside in the intergalactic medium (IGM), where the extremely low densities ($n_\mathrm{H}\sim 10^{-6} \mathrm{cm}^{-3}$) and high temperatures ($T\sim 10^6~\mathrm{K}$) make observational confirmation elusive due to the lack of UV and optical transition lines in this temperature-density regime \citep{Cen1999, Bregman2007, Shull2012}.




Fast Radio Bursts (FRBs) offer a promising tool to find these missing baryons by directly probing the ionised gas that has eluded the emission and absorption line census to date: highly ionised low density material \citep{McQuinn2014, Deng2014, Keane2019, Macquart2020}. FRBs are a class of newly discovered, bright extragalactic radio transients ($\sim1~\mathrm{Jy}$) with short, millisecond durations and unknown origins \citep{Lorimer2007, Thornton2013}. For recent reviews on FRBs see \citet{Petroff2019} and \citet{Cordes2019}.

The majority of the 137 FRBs detected to date (November 2020) have been one-off events -- only 22 have been seen to repeat \citep{Spitler2014, Spitler2016, Petroff2016, Marcote2017, CHIME2019_second_repeater, CHIME2019_Repeaters}. Currently, the host galaxies have been identified for three repeating (FRB~21102; \citealt{Tendulkar2017}, FRB~180916; \citealt{Marcote2020}, FRB 190711; \citealt{Macquart2020}) and nine non-repeating FRBs (FRB~180924; \citealt{Bannister2019}, FRB~190523; \citealt{Ravi2019}, FRB~181112; \citealt{Prochaska2019b}, FRB~190102, FRB~190611, FRB~190711; \citealt{Macquart2020}, FRB~191001; \citealt{Bhandari2020}, FRB~190717 and FRB~200430; \citealt{Heintz2020}).



One of the defining features of FRBs, and the key to finding the missing baryons, is their large dispersion measures (DM\footnote{We note that in this work we are using convention in the FRB literature of using the initialism `DM' to stand for `dispersion measure' instead of `dark matter' as is the convention in cosmology and hydrodynamic simulations. We will be explicitly using `dark matter' in the text when required.}) relative to the DM of the interstellar medium. 

The DM of an FRB is an observed quantity that measures the delay in arrival time of the burst as a function of frequency. Electromagnetic waves travelling in an ionised medium will experience a frequency dependent delay. The lower frequency electromagnetic waves of the FRB will be delayed relative to higher frequencies, causing the pulse to become `dispersed'. The larger the quantity of plasma along the line-of-sight, the larger the observed time delay between frequencies an (i.e the larger the DM). 

The observed DMs of FRBs currently span more than an order of magnitude: the lowest FRB DM is 103~$\pccm$ \citep{CHIME2019_Repeaters} and the largest is 2596~$\pccm$ \citep{Bhandari2018}. 

Taking into account cosmological effects, the DM of a source at redshift $z$ is exactly equal to:

\begin{equation}
    \mathrm{DM} = \int_0^z \frac{n_e(z)}{1 + z}\,\mathrm{d}l\,,
    \label{eq:dispersion_measure}
\end{equation}

\noindent where $n_e$ is the physical electron density, $\mathrm{d}l$ is the physical distance element such that $\mathrm{d}l = c(1+z)^{-1} \mathrm{H}_0^{-1} E(z)^{-1} \mathrm{d}z$, $E(z)~=~\sqrt{\Omega_m (1+z)^3 + \Omega_{\Lambda}}$, $c$ is the speed of light, $\mathrm{H}_0$ is the Hubble parameter at redshift $z = 0$, $\Omega_b$ and $\Omega_{\Lambda}$ are the cosmic matter density and cosmic dark energy density respectively. The $(1 + z)$ factor in the denominator accounts for cosmological time dilation due to the expanding Universe.

Because FRBs originate from galaxies external to the Milky Way, it is convenient to break the observed DM ($\mathrm{DM_{Obs}}$) into 3 components: $\mathrm{DM_{MW}}$, $\mathrm{DM_\mathrm{cosmic}}(z)$ and $\mathrm{DM_{Host}}$ as shown in \Cref{eq:DM_components}.
\begin{equation}
    \mathrm{DM_{Obs}}(z) = \mathrm{DM_{MW}} + \mathrm{DM_{cosmic}}(z) + \frac{\mathrm{DM_{Host}}}{1+z} \,,
    \label{eq:DM_components}
\end{equation}
where $\mathrm{DM_{MW}}$ is the DM due to the the Milky Way (this includes the ionised plasma in both the interstellar medium and the circumgalactic medium), $\mathrm{DM_\mathrm{cosmic}}(z)$ is the DM component due to the IGM and any contributions from intervening galaxy halos along the line-of-sight, and $\mathrm{DM_{Host}}$ is the DM component due to the host galaxy and local source environment of the FRB. 

The $\mathrm{DM_{MW}}$ component is usually estimated using a galaxy electron density model such as NE2001 \citep{NE2001} or YWM16 \citep{YMW2016}. This quantity is typically a small fraction of $\mathrm{DM_{Obs}}$. For example, FRB 121002 has a $\mathrm{DM_{Obs}}$ of 1629.18 $\pccm$, whereas the NE2001 and YMW16 models estimate a $\mathrm{DM_{MW}}$ in the direction of the burst of $72.2~\pccm$ and $60.5~\pccm$ respectively. It should be noted that these models have their own uncertainties of a factor 2 or 3 (Price et al. \textit{submitted}).

The $\mathrm{DM_{Host}}$ is currently the least constrained parameter of the three contributing components of $\mathrm{DM_{Obs}}$. The value of $\mathrm{DM_{Host}}$ is unique for each galaxy, modulated further by the path of the FRB with respect to the plane of the galaxy. For example, a highly inclined galaxy is likely to have a higher $\mathrm{DM_{Host}}$ because the burst must traverse the ionised material in the denser ISM in the disk of the galaxy. Additionally there may be some contributions to $\mathrm{DM_{Host}}$ from the local source environment around the FRB. There have been some attempts to provide estimates of $\mathrm{DM_{Host}}$ \citep{Yang2017}, however it is frequently assumed $\mathrm{DM_{Host}} = 0$ or a value similar to the Milky Way ($\mathrm{DM_{Host}} = 100~\mathrm{pc~cm^{-3}}$).

The third component, $\mathrm{DM_\mathrm{cosmic}}$, is the focus of this paper. The $\mathrm{DM_\mathrm{cosmic}}$ component includes both the contribution from the IGM ($\mathrm{DM_{IGM}}$) and any intervening galaxy halos ($\mathrm{DM_{Halo,Int.}}$). Whilst $\mathrm{DM_{Halo,Int.}}$ is unique to the line-of-sight towards each FRB, $\mathrm{DM_{IGM}}$ can be analytically estimated using \Cref{eq:DM_IGM} as derived in \citet{Deng2014}:

\begin{equation}
    \mathrm{DM_{IGM}} = \frac{3 c \mathrm{H}_0 \Omega_b}{8 \pi G m_p} \large\int_0^z \frac{ (1 + z)f_{\mathrm{IGM}}(z) \chi (z)}{\sqrt{\Omega_m (1+z)^3 + \Omega_\Lambda}} dz  \,,
    \label{eq:DM_IGM}
\end{equation}
where $\Omega_b$ is the cosmic baryon density, $G$ is the gravitational constant, $m_p$ is the proton mass and $f_\mathrm{IGM}(z)$ is the fraction of baryons in the IGM at redshift $z$. For a model of the IGM containing only hydrogen and helium, the ionisation parameter $\chi(z)$ is given by:

\begin{equation}
    \chi (z) = \left[0.75 y_1 \chi_{e, \mathrm{H}}(z) + 0.25 \frac{y_2}{2} \chi_{e,\mathrm{He}} (z)\right]  \,,
    \label{eq:ionisation_parameter}
\end{equation}
where $\chi_{e, \mathrm{H}}(z)$ and $\chi_{e, \mathrm{He}}(z)$ are the ionisation fractions of hydrogen and helium respectively, and $y_1 \sim 1$ and $y_2 \simeq 4 - 3y_1 \sim 1$ indicating the small deviations from the $0.75$-to-$0.25$ split of primordial hydrogen and helium mass fractions. The helium ionisation factor has an additional factor of $\frac{1}{2}$ due to helium only providing 0.5 electrons per proton mass.

The key to identifying the missing baryons is the relationship between the DM and the redshift (i.e the $\mathrm{DM} - z$ relation, also the Macquart $\mathrm{DM} - z$ relation) of FRBs. The shape and slope of the $\mathrm{DM} - z$ relation depends on the quantity of ionised baryons in the IGM. Therefore by measuring the host galaxy redshift and the $\mathrm{DM_{IGM}}$ component of FRBs and comparing with a theoretical $\mathrm{DM} - z$ relation, it becomes possible to measure the fraction of baryons in the IGM. This exact analysis was performed by \citet{Macquart2020} to conclude that FRBs are able to find all the `missing baryons' in the IGM.


The $\mathrm{DM} - z$ relations in the literature have been established using various techniques including: analytic models, semi-analytic models, and hydrodynamic simulations. Each of these techniques have significant differences in both the assumptions that they make and the resulting $\mathrm{DM} - z$ relation.

The analytic models of \citet{Ioka2003}, \citet{Inoue2004} and \citet{Zhang2018} approximate the $\mathrm{DM} - z$ relation through the cosmic baryon density, $\Omega_b$, in addition to assuming a composition and ionisation state of the Universe. The differences in the analytic models arise due to these differing assumptions.

The \citet{Ioka2003} model matches a homogeneous Universe with all the baryons ($\Omega_b = 0.044$) fully ionised. It also assumes the Universe is homogeneously filled with 100\% ionised hydrogen. These assumptions lead to an approximately linear $\mathrm{DM} - z$ relation with a slope of $\mathrm{DM_{\mathrm{Ioka}}}(z)/z~\approx~1174~\pccm$.

The \citet{Inoue2004} and \citet{Zhang2018} models are is similar to \citet{Ioka2003}, except with different parameters. \citet{Inoue2004} instead assumes the IGM consists of 24\% helium by mass and the rest hydrogen. Additionally it models hydrogen as fully ionised and helium as singly ionised. The \citet{Zhang2018} model on the other hand adds two additional parameters: the helium abundance and the baryonic fraction locked inside galaxies. \citet{Zhang2018} assumes that all baryons are fully ionised and a 0.875-to-1 ratio between electrons and baryons (to account for helium) and also assumes that only 85\% of baryons are in the IGM (with the remaining 15\% locked up within galaxies and the CGM). These differing assumptions lead to a flatter $\mathrm{DM} - z$ slope of $\mathrm{DM_{\mathrm{Inoue}}}(z)/z~\approx~960~\pccm$ and $\mathrm{DM_{\mathrm{Zhang}}}(z)/z~\approx~850~\pccm$ for \citet{Inoue2004} and \citet{Zhang2018} respectively. 

Since the analytic formulations all assume a homogeneous Universe, there is no scatter around the relation. These analytic models can not provide information on the line-of-sight variations due to the clumpy IGM. There has been additional work to calculate analytic estimates of the scatter around the $\mathrm{DM} - z$ relation by \citet{McQuinn2014} and \citet{Macquart2020}. However, simulations are required to support these estimates of line-of-sight variations.

The semi-analytic approach from \citet{Pol2019} uses a large $N$-body/dark matter simulations to estimate the electrons density in the IGM from the dark matter distribution. This technique allows the use of extremely large simulations with high-resolution, however the drawback is that it only use dark matter density distributions and do not include any baryonic physics.

\citet{Pol2019} used the MareNostrum MICE large $N$-body/dark matter simulations \citep{Fosalba2008} which have box sizes $3072\ \mathrm{Mpc}\ h^{-1}$ with $2048^3$ particles. These simulations are a series of concentric radial shells of finite width around a central observer. 

\citet{Pol2019} convert the dark matter density into a free electron density and then integrate this to find the DM as a function of redshift. They find $\mathrm{DM_{cosmic}}$ at redshift $z=1$ to be $\mathrm{DM_{cosmic}}(z=1)~=~800^{+7000}_{-170}~\mathrm{pc~cm^{-3}}$. However, their $\mathrm{DM} - z$ relation has a significantly reduced estimate of IGM baryons compared to all other studies at low redshifts ($z < 1$). Their 95\% confidence interval predicts IGM column densities of electrons equivalent to that of the Milky Way halo ($\sim 30-100\ \pccm$) out to redshift $z=0.5$.

Studies of the $\mathrm{DM} - z$ relation using cosmological hydrodynamic simulations have previously been used by \citet{McQuinn2014}, \citet{Dolag2015} and \citet{Jaroszynski2019}. Each of these have used a different suite of simulations and arrive at different estimates of the $\mathrm{DM} - z$ relation. 

\citet{McQuinn2014} calculated the $\mathrm{DM} - z$ relation with different models (analytic and simulations) for the distribution of cosmic baryons. These showed that the variance around the $\mathrm{DM} - z$ relation is quite sensitive to whether the `missing' baryons are at the boundary of the IGM and CGM or significantly further out. From their cosmological simulation they found a variance around the mean $\mathrm{DM} - z$ relation of $40\%$. However their simulation was relatively small (40 $h^{-1}$ cMpc; For comparison: \EAGLE\ = 100 cMpc, Magneticum Pathfinder = 896 $h^{-1}$ cMpc and Illustris = 75 $h^{-1}$ cMpc)  and do not specify the number of lines-of-sight they have used.

\citet{Dolag2015} used the Magneticum suite of simulations to constrain the origins of FRBs by calculating the DM contributions from the Milky Way, local Universe and large scale structure. The Magneticum Pathfinder simulations are large enough (896 $h^-1$ Mpc) that their simulation boxes overlap in redshift, however their resolution is significantly lower than \EAGLE\ and Illustris. They used $4096^2~(\sim 1.7\times 10^6)$ sight lines to generate the DM probability distribution function at 7 redshifts between $0 < z < 1.980$.

\citet{Jaroszynski2019} used the Illustris simulations and found a $13$ percent scatter around the $\mathrm{DM} - z$ relation at $z~=~1$ and $7$ percent at $z~=~3$. They also simulated populations of FRBs to determine if they are able constrain cosmological parameters such as the Hubble constant, $\mathrm{H}_0$.
They concluded that to constrain the mean ionised fraction to $\sim 1\%$ at various epoch, would require $10^4$ FRBs with known redshifts, a number significantly higher than the current number of measured FRB redshifts.

In this work we have used the \EAGLE\ simulations to calculate the $\mathrm{DM} - z$ relation from redshift $z = 0$ to 3 using over a billion lines of sight. This is the largest number of lines of sight used in the analysis of the $\mathrm{DM} - z$  relation; exceeding the previous highest by \citet{Jaroszynski2019} by a factor of 4. In doing so we have produced $\mathrm{DM_{cosmic}}$ PDFs at the highest number of redshift samples (66 samples between redshifts $z = 0$ and 3).
We compare our results to those in the literature and find that the mean $\mathrm{DM_{cosmic}}$ broadly agrees with previous work and is well fitted by a non-linear model that includes parameters for cosmology.
We have performed an explicit investigation into the scatter around the mean finding that there is significant asymmetry in the shape of the PDFs. We also provide convergence tests for both volume and resolution, which has previously not been performed.
We also compare our $\mathrm{DM} - z$ relation to the redshifts of known FRB host galaxies and find that many FRBs lay in the $2 - 3 \sigma$ region of the confidence interval. We suggest that this could be an indication that these FRBs have intersected with filaments of the IGM.

In \Cref{sec:EAGLE} we provide an overview of the \EAGLE\ cosmological hydrodynamic simulations. In \Cref{sec:creating_dm_maps} we describe the process of generating projected DM maps from \EAGLE. In \Cref{sec:interpolated_dm_maps} we describe the process of generating interpolated DM maps for 66 redshift samples and the process of combining these maps to produce a $\mathrm{DM} - z$ relation in \Cref{sec:cumulative_sum}. In \Cref{sec:Results} we present the DM-redshift relation and discuss the scatter around the mean. We also compare with observations of localised FRBs. Finally, we summarise our results and provide our conclusions in \Cref{sec:Conclusions}.


Length units with the prefixes `p' and `c' indicate `proper' and `comoving' quantities respectively. The exception is centimetres (cm) which is in proper units.

\section{The EAGLE Simulations}\label{sec:EAGLE}
\begin{table*} 
\caption{The simulation properties and lines-of-sight parameters used in this work. From left to right the columns are: simulation name, comoving box size ($L_\mathrm{box}$), number of dark matter particles ($N$; this is the same as the initial number of gas particles), the resolution level of the simulation, initial mass of the baryonic particles ($m_\mathrm{gas}$), mass of the cold dark matter particles ($m_\mathrm{cdm}$), maximum physical softening length ($\epsilon_\mathrm{phys}$), the number of pixels ($N_\mathrm{pixels}$; equivalent to the number of lines-of-sight) per DM map, the number of DM maps generated ($N_\mathrm{z}$) and the convergence test associated with the simulation. The name of each simulation consists of three sections:\texttt{<prefix>L<box size>N<particles>}. The prefix indicates the stellar/AGN feedback model used (\texttt{Ref} indicating a reference simulation and \texttt{Recal} having the stellar and AGN feedback fine tuned to better fit the galaxy stellar mass function at high resolution). Columns 2-7 are properties of the simulation themselves; columns 8-9 are parameters chosen during our analysis. The main simulation we refer to in our results is RefL0100N1504 with the remaining used to perform convergence tests.}
\begin{tabular}{lcccccccccl}
\hline
Name & $L_\mathrm{box}$ & $N$ & Resolution & $m_{\mathrm{gas}}$ & $m_{\mathrm{cdm}}$ & $\epsilon_{\mathrm{phys}}$ & $N_\mathrm{pixels}$ & $N_z$ & Associated  \\ 

& $(\mathrm{cMpc})$ & {} & {} & $\left(\mathrm{M}_{\odot}\right)$ & $\left(\mathrm{M}_{\odot}\right)$ & $(\text {pkpc})$ & & & Convergence Test \\ 
\hline
\hline
RefL0100N1504 & 100 & $1504^3$ & Medium & $1.81 \times 10^6$ & $9.70 \times 10^6$ & 0.70 & $32000^2$ & 66 & Reference \\
RefL0050N0752 & 50 & $752^3$ & Medium & $1.81\times 10^6$ & $9.70 \times 10^6$ & 0.70 & $16000^2$ & 131 & Volume\\
RefL0025N0376 & 25 & $376^3$ & Medium  & $1.81 \times 10^6$ & $9.70 \times 10^6$ & 0.70 & $8000^2$ & 262 & Volume and Resolution\\ 
RefL0025N0752 & 25 & $752^3$ & High  & $2.26 \times 10^5$ & $1.21 \times 10^6$ & 0.35 & $8000^2$ & 262 & Resolution and Physics\\
RecalL0025N0752 & 25 & $752^3$ & High  & $2.26 \times 10^5$ & $1.21 \times 10^6$ & 0.35  & $8000^2$ & 262 & Physics\\
\hline

\end{tabular}
\label{tab:EAGLE_sim}
\end{table*}

In this section we provide an overview of the \EAGLE\ (Evolution and Assembly of GaLaxies and their Environments) simulations. For a more detailed description of the \EAGLE\ simulations see \cite{Schaye2015} and \cite{Crain2015}. \par

The \EAGLE\ simulations are a suite of high resolution cosmological $N$-body/hydrodynamic simulations performed using a modified version of the smooth particle hydrodynamic (SPH) code \textsc{gadget-3}, last described in \cite{Springel2005}. 
\EAGLE\ adopts a \citet{Planck2014} $\Lambda$CDM cosmology, with the following parameters: $\Omega_m = 0.307$, $\Omega_\Lambda = 0.693$, $\Omega_b = 0.04825$, $H_0 = 67.77~\mathrm{km~s^{-1}~Mpc^{-1}}$, $\sigma_8 = 0.8288$, $n_s = 0.9611$ and $Y = 0.248$.

The main \EAGLE\ simulation cubes were run with volumes of 25, 50 and 100 cMpc per side and employed a gravitational softening length of 0.7 pkpc. The resolution in \EAGLE\ is not sufficient to fully resolve the multi-phase nature of the interstellar medium (ISM), but is able to marginally resolve the warm ISM ($\mathrm{T \sim 10^4~K}$) on $\sim \mathrm{kpc}$ scales.

Galactic-scale processes that occur on scales that are unresolved in \EAGLE\ (these processes are often referred to as `subgrid physics') are implemented using subgrid models. 
Gas is cooling is implemented through tracking the abundances of 11 elements (H, He, C, N, O, Ne, Mg, Si, S, Ca, and Fe) and their cooling rates following \cite{Wiersma2009a}.
Hydrogen reionisation is implemented by `turning-on' a \cite{HaardtMadau2001} time-dependent spatially uniform UV/X-ray background at $z = 11.5$. 
Star formation follows \cite{Schaye2008} and, by-design, reproduces the Kennicutt-Schmidt star formation law \citep{Kennicutt1998}.
Star particles are treated as simple stellar populations with a \cite{Chabrier2003} IMF with stellar evolution and mass loss based on \cite{Wiersma2009b}.
Galactic winds are driven by energy feedback from star formation \citep{DallaVecchia2012} and active galactic nuclei (AGN; \citealt{Booth2009}) through the stochastic heating of gas particles to temperatures large enough to overcome the over-cooling problem.

We have chosen to use the \EAGLE\ simulations to study the $\mathrm{DM} - z$ relation of the IGM for the following reasons:

\begin{enumerate}
\item The physics scheme implemented in \EAGLE\ is successful in reproducing observed galaxy properties at low redshift such as the cosmic star formation history~\citep[e.g.][]{Madau1996} in \citet{Furlong2015}.
\item \citet{Rahmati2015} showed that \EAGLE\ is in broad agreement with observed \HI\ absorption line statistics \citep{Rudie2012, Prochaska2013}. Since \HI\ absorption and DM are linked (\HI\ absorption traces neutral gas, whereas DM traces ionised gas), it should be expected that \EAGLE\ is also well suited to predicting the $\mathrm{DM} - z$ relation of the IGM. 
\item The resolution in \EAGLE\ is high enough to resolve the Jeans length in the IGM.
\item  Since \EAGLE\ is a hydrodynamic simulation, it evolves both the baryonic matter and dark matter together self-consistently. Hence we do not make any assumptions about the distribution of baryons based on dark matter distributions, as is done with semi-analytic models.
\end{enumerate}

\Cref{tab:EAGLE_sim} summarises the simulations that were used in this work. The reference feedback model was calibrated at the standard \EAGLE\ resolution (i.e the resolution of RefL0100N1504). The reference feedback model is used in the simulations listed in \Cref{tab:EAGLE_sim} with the prefix `Ref'. The simulation RecalL0025N0752 was re-calibrated in the same manner as the reference feedback model except at eight times higher resolution. Using a variety of box sizes (i.e. 25, 50 and 100 cMpc), resolutions and feedback calibrations is necessary to test for resolution and box size convergence. Except for in \Cref{app:convergence_tests,,app:shuffled_vs_transformed} we will only be sharing the results obtained from the analysis of RefL0100N1504.

\section{Methods} \label{sec:methods}
We have used \EAGLE\ to calculate the $\mathrm{DM} - z$ relation between redshifts $0 < z \lesssim 3$. We limit our analysis to redshifts $z \lesssim 3$ because helium reionisation is expected to be complete by redshift $z \sim 3$ and the \cite{HaardtMadau2001} UV background may not accurately reproduce the abundances of singly ionised helium (i.e. at redshift $0 < z \lesssim 3$ the properties of the IGM are relatively simple).

\subsection{Creating Dispersion Measure Maps} \label{sec:creating_dm_maps}
For each of the simulations listed in \Cref{tab:EAGLE_sim}, we produced integrated electron column density ($N_e$) maps from all the \EAGLE\ snapshots in the redshift range $z = 0$ to $3.016$. We then converted each of these column density maps into DM maps.

The method we used to calculate these integrated column densities is detailed in Section 2.2 of \cite{Wijers2019}; however we provide an overview of this process in \Cref{sec:cd_calculation} below.



\subsubsection{Column Density Calculation}\label{sec:cd_calculation}
We post-processed the \EAGLE\ snapshots to obtain the ion abundances of \HI, \HeI\ and \HeII. We determined the number density of ions using tabulated ionisation fractions as a function of density, temperature, redshift. These ionisation fraction tables were computed using the spectral synthesis program Cloudy (version c07.02.00, \citealt{Ferland1998}) under the same cooling assumptions that were used in \EAGLE. We used the \cite{Rahmati2013} prescription to obtain the fraction of neutral hydrogen because the ionisation tables do not account for self-shielding against ionising radiation in high-density ($n_\mathrm{H} \gtrsim 10^{-3} ~\mathrm{cm}^{3}$) gas.

We obtained the column densities ($N_x$; where $x$ is H, \HI, He, \HeI\ or \HeII) of these ions by summing within columns (thin elongated rectangular prisms) of fixed length and finite area.
We divided the $X$ and $Y$ directions of a snapshot such that each column had an area of $3.125^2~\mathrm{ckpc}^2$. This area was chosen such that the column density statistics had converged. The length of each column is equal to the box size ($L_\mathrm{box}$) This gives us the number of lines-of-sight ($N_\mathrm{pixels}$) shown in \Cref{tab:EAGLE_sim}.

We use the SPH kernel to project the ion abundances within SPH particles into columns, then add the particle contributions in each column together. We then divide the total abundance in each column by the area ($3.125^2~\mathrm{ckpc}^2$) to get the column densities in lines-of-sight along the $Z$-axis.\footnote{We only used projections along the $Z$ axis instead of three axes, because performing projections along $X$, $Y$ and $Z$ would triple the computation time.} We refer to this 2D projection as a column density `map'. Each pixel in the map is the column density along that line-of-sight.

\subsubsection{Electron Column Density Maps}\label{sec:electron_cd_maps}
We calculated the electron column density, $N_{e}$, along each line-of-sight from the calculated ion column densities as shown in \Cref{eq:electron_column_density}, 
\begin{align}
    N_e = \left(N_{\mathrm{H}} - N_{\HI}  \right) + \left(2 N_{\mathrm{He}} - 2N_{\HeI} - N_{\HeII} \right).
    \label{eq:electron_column_density}
\end{align}

Here $N_{\mathrm{H}}$, $N_{\mathrm{H{\textsc{i}}}}$, $N_{\mathrm{He}}$, $N_{\mathrm{He{\textsc{i}}}}$, $N_{\mathrm{He{\textsc{ii}}}}$ are the column densities of hydrogen, neutral hydrogen, helium, neutral helium and singly ionised helium respectively. The values for $N_e$ were determined in this manner because we did not have ionisation tables to calculate ion fractions for singly ionised hydrogen (\HII) and doubly ionised helium (\HeIII). 

We converted these electron column density maps into DM maps by a change of units (from $\mathrm{cm^{-2}}$ to $\mathrm{pc~cm^{-3}}$). We note that for snapshots with redshift $z > 0$, this is not the `true' DM map at that redshift, but the DM map of that snapshot as if it was at redshift $z = 0$. We corrected for this by introducing a factor of $(1+z)^{-1}$ (see \Cref{sec:interpolated_dm_maps} and \Cref{app:dm_derivation} for a derivation of this factor). In \Cref{fig:DM_Maps_3_panel} we show three of the DM maps at redshifts $z = 0, 1.004$ and $3.017$.

We clarify here that in this step we are not computing the total DM to a source, but the DM over the length of the \EAGLE\ boxes at different redshifts. 

In \Cref{sec:interpolated_dm_maps} we generate interpolated DM maps at redshift intervals equal to the $L_\mathrm{box}$. We sum these interpolated DM maps in \Cref{sec:cumulative_sum} to obtain the total DM to sources at different redshifts.

\subsection{Interpolated Dispersion Measure Maps}
\label{sec:interpolated_dm_maps}
To study the $\mathrm{DM} - z$ relation for distances larger than one box size, we need to connect snapshots together (as discussed above) by creating continuous lines-of-sight from redshifts $z = 0$ to 3. However the \EAGLE\ boxes do not overlap because the redshift spacing between the snapshots is larger than the box size ($L_\mathrm{box}$). In other words, there are `gaps' in redshift that \EAGLE\ does not cover. Without creating maps to fill in these redshift `gaps' we would be significantly underestimating the line-of-sight DM.

For each of the simulations we generated a sequence of sample redshifts between $z = 0$ and $3.016$ (the redshift of \EAGLE\ snapshot 17) separated by $L_\mathrm{box}$. 
The redshift samples were generated assuming the same cosmology as implemented as in \EAGLE\ \citep{Planck2014}.
This gave us $N_z$ redshift samples (see \Cref{tab:EAGLE_sim} for a summary of the simulations used).

We calculated an interpolated DM map for each of the redshift samples by linearly interpolating between neighbouring \EAGLE\ DM maps.
In other words; to create an interpolated DM map at redshift $z = 0.2$ in the RefL0100N1504 simulation, we linearly interpolated between the DM maps created from neighbouring snapshot outputs at redshifts $z = 0.18$ and  $z = 0.27$.

We linearly interpolated in redshift rather than using comoving distance because as shown in \Cref{eq:DM_derivation_5} DM scales linearly with the scale factor $a = (1 + z)^{-1}$.
We multiplied each of these interpolated DM maps by a factor of $(1 + z)^{-1}$ to obtain the true DM values at that redshift (see \Cref{app:dm_derivation} for a derivation of this $(1+z)^{-1}$ factor).

\begin{figure*}
    \centering
    \includegraphics[clip, width=\linewidth]{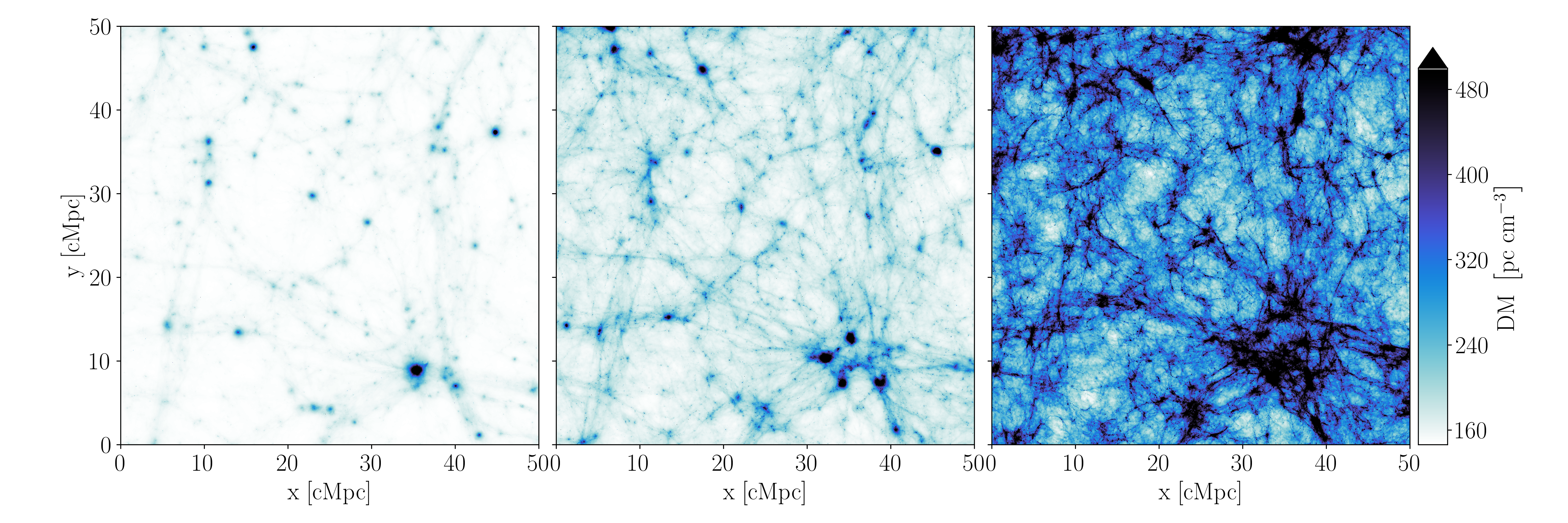}
    \caption{Left to Right: Zoom in DM maps from simulation RefL0100N1504 at several redshifts ($z= 0, 1.004$ and $3.017$, from left to right). Each of these maps span 50 cMpc $\times$ 50 cMpc (i.e. one quadrant of the larger DM map) with $16,000 \times 16,000$ lines-of-sight. The larger DM values in the third panel compared to the first panel are not due to longer lines-of-sight (these maps only span the 100 cMpc simulation box size) but due to the average physical density of the Universe being larger at higher redshifts. Note that we have capped the colour bar at 500 $\mathrm{pc\ cm^{-3}}$ to emphasise the filament structure of the IGM.}
    \label{fig:DM_Maps_3_panel}
\end{figure*}

This gave us $N_z$ DM maps, all separated spatially by $L_\mathrm{box}$. Because the spatial separation between the maps is $L_\mathrm{box}$, we were able to add these maps together to obtain a continuous sight lines up to redshift $z = 3.017$. In \Cref{fig:DM_Maps_3_panel} we show a quadrant from three DM maps at redshifts $z = 0, 1.004$ and $3.017$.

\subsubsection{Pixel Randomisation (`Scramble Technique' )}\label{sec:shuffle}
A common issue that arises when connecting simulation boxes to create larger volumes is the replication of structure. If these boxes were naively combined as is, then there would be a periodic repeating of structure along each line-of-sight. This would lead to results that would not match our Universe.

One way to minimise the effects of periodic repeating of structure is to perform a randomised transformation on each replicated simulation box \citep{Blaizot2005}. These transformations are a series of rotations, mirrors and translations. 

However, as we show in \Cref{app:shuffled_vs_transformed}, in the case of DM maps this transformation technique is ineffective at removing correlations in the smaller simulations (25 and 50 cMpc). This is because there is less space in the smaller boxes to `translate' the boxes, which leads to a greater number of repeating structures. 

We have instead used a `scramble' technique to remove the structure of the DM maps. For each map we randomly reassign the individual lines-of-sight (pixels) to new locations. The reassignment of lines-of-sight to new position is completely independent for each map. This process scrambles the positions of the lines-of-sight to ensure there we have no repeating structures.

\subsection{Cumulative Sum and PDF Normalisation} \label{sec:cumulative_sum}
After scrambling the positions of the lines-of-sight we have a 3-dimensional array ($N_\mathrm{pixels} \times N_\mathrm{pixels} \times N_z$) containing DM maps separated by $100~\mathrm{cMpc}$ with no periodic structure (see \Cref{sec:shuffle}). 

For each pixel, we performed a cumulative sum along $z$. This gave us $N_\mathrm{z}$ redshift samples, each with $N_\mathrm{pixels} \times N_\mathrm{pixels}$ line-of-sight DM measurements.

For each of the $N_\mathrm{z}$ redshift samples we constructed the DM probability density function (PDF). We created histograms using 1000 ($N_\mathrm{bins}$) logarithmically spaced bins between $10^0 - 10^5\ \mathrm{pc\ cm^{-3}}$. We normalised each histogram to produce the probability density function (PDF). This normalisation is:


\begin{equation}
    \mathrm{PDF}(z) = N_{\mathrm{los}, i} \left(\Delta \mathrm{DM}_i\sum_{i=0}^{N_\mathrm{bins}} N_{\mathrm{los}, i}\right)^{-1} \,,
    \label{eq:normalisation}
\end{equation}
where $\mathrm{PDF}(z)$ is the $\mathrm{DM}_\mathrm{cosmic}$ PDF at redshift $z$, $N_{\mathrm{los}, i}$ is the number of lines-of-sight within the $i$th DM bin, and $\Delta \mathrm{DM}_i$ is the width of the $i$th DM bin.

\section{Results and Discussion}\label{sec:Results}

\subsection{Dispersion Measure - Redshift Relation}
\label{sec:dmz-relation}

\begin{figure*}
    \centering
    \includegraphics[width=\linewidth]{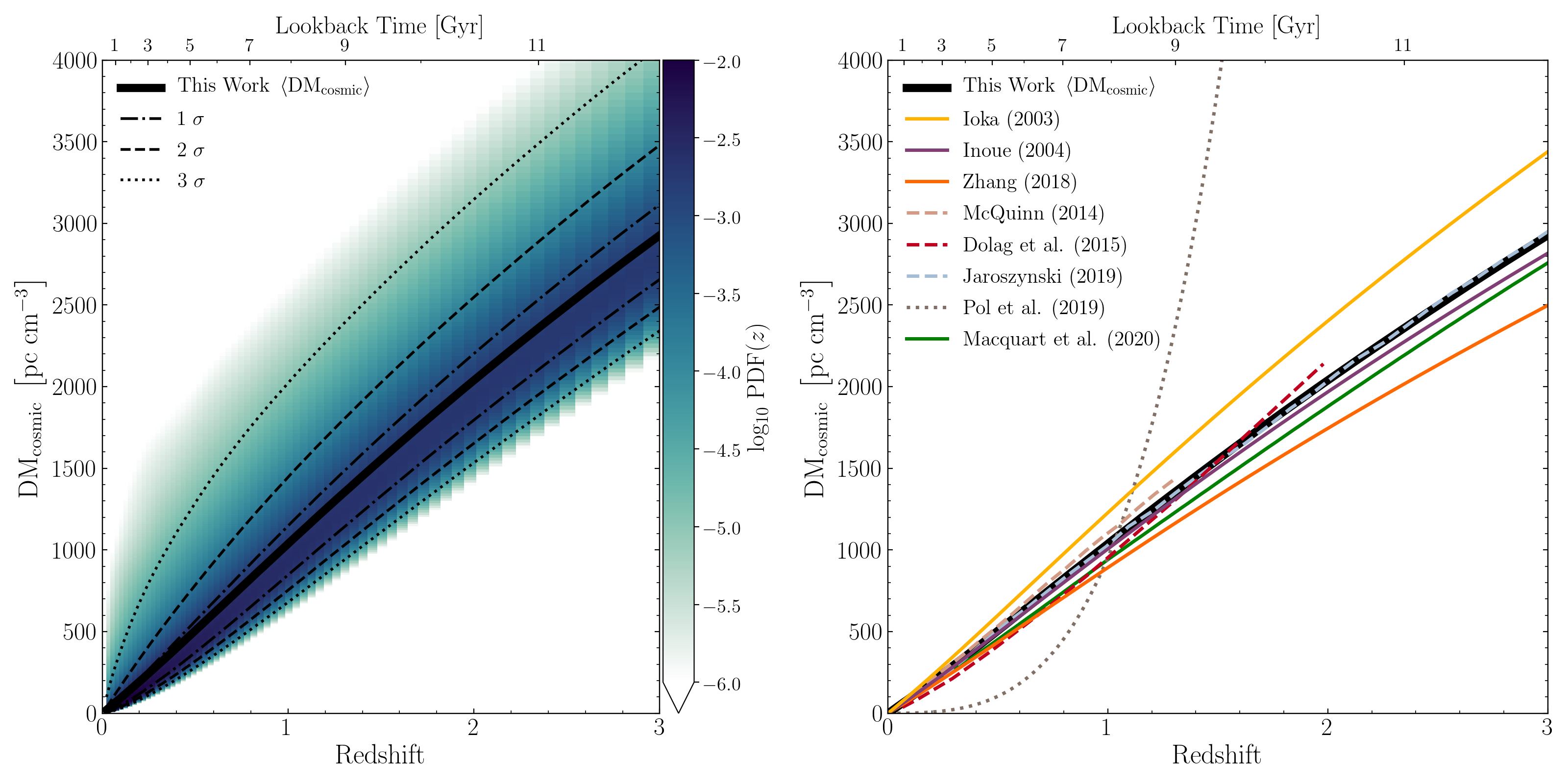}
    \caption{\emph{Left:} The purple-green coloured histogram is the PDF of \DMcosmic\ in the \mbox{EAGLE} RefL0100N1504 simulation at each redshift. The thick solid black line is the mean \DMcosmic\ at each redshift. The thin black lines are the $1~\sigma$ (dash-dotted), $2~\sigma$ (dashed) and $3~\sigma$ (dotted) confidence intervals at each redshift. The \DMcosmic\ PDF at has been normalised to unity at each redshift.
    \emph{Right:} A comparison between the mean $\mathrm{DM} - z$ relation of this work and other results in the literature. The thick solid black line is the mean \DMcosmic\ at each redshift (same as left). The remaining solid lines are the analytic models of \citet{Ioka2003} (yellow), \citet{Inoue2004} (purple), \citet{Zhang2018} (orange) and \citet{Macquart2020} (green). These analytical approximations have zero spread in DM values due to assuming the universe is homogeneous. The differences in these analytical formulations are due to differing assumptions of the elemental composition, ionisation state and baryonic content of the IGM. The dashed lines are hydrodynamic models of \citet{McQuinn2014} (light brown), \citet{Dolag2015} (red), and \citet{Jaroszynski2019} (light-blue). The dotted line is the semi-analytic model of \citet{Pol2019} (grey).}
    \label{fig:DMz_Hist}
\end{figure*}

In the left panel of \Cref{fig:DMz_Hist} we show the PDFs of \DMcosmic\ from the RefL0100N1504 simulation at 66 redshift intervals between $0~<~z~<~3.016$. The thick solid black line is the mean \DMcosmic\ ($\left\langle \mathrm{DM_{cosmic}} \right\rangle$) at each redshift and the dot-dashed, dashed, and dotted lines are the $1\sigma$, $2\sigma$, and $3\sigma$, confidence intervals respectively. See \Cref{sec:mean_dm} and \Cref{sec:sigma_dm} for more details on $\left\langle \mathrm{DM_{cosmic}} \right\rangle$ and the confidence intervals. The PDFs have been normalised to unity as described in \Cref{eq:normalisation}. We have truncated the colour bar at $\log_{10} \mathrm{PDF}(z) = -6$ because below this value there is noise in the PDF bins due to the low number of lines-of-sight. 

In the right panel of \Cref{fig:DMz_Hist}, we plot the $\left\langle \mathrm{DM_{cosmic}} \right\rangle$ with existing analytic (solid), hydrodynamic (dashed) and semi-analytic (dotted) analysis in the literature. The analytic models of \citet{Ioka2003}, \citet{Inoue2004}, \citet{Zhang2018} and \citet{Macquart2020} are coloured yellow, purple, orange and green respectively. The hydrodynamic models of \citet{McQuinn2014}, \citet{Dolag2015} and \citet{Jaroszynski2019} are coloured light brown, red and light-blue respectively. The semi-analytic model of \citet{Pol2019} is coloured grey. The line for \citet{Jaroszynski2019} is difficult to see because it almost perfectly overlaps with $\left\langle \mathrm{DM_{cosmic}} \right\rangle$. 
We find a large spread in the \DMcosmic\ PDFs. This is not unexpected since this spread is indicative of the variation in cosmic electron column density between different lines-of-sight. See \Cref{sec:sigma_dm} for further analysis of the non-Gaussianity of these PDFs.

Our $\left\langle \mathrm{DM_{cosmic}} \right\rangle$ broadly agree with the other models in the literature; in-particular \citet{Jaroszynski2019} is an extremely close match for all redshifts.
   
The \citet{Ioka2003} can be considered an upper-limit to the slope of the $\mathrm{DM}-z$ relation because it assumes the Universe is homogeneously filled with ionised hydrogen alone. Increasing the helium fraction or the amount of baryons locked inside galaxies both decrease the slope of the $\mathrm{DM} - z$ relation.

On the other hand, \citet{Zhang2018} uses a $f_\mathrm{IGM}$ factor ($f_\mathrm{IGM} = 0.85$) to analytically exclude baryons locked inside galaxies from contributing to $\mathrm{DM_{cosmic}}$. However, their model underestimates $\left\langle \mathrm{DM_{cosmic}}\right\rangle$ at all redshifts compared to the results of this work and other hydrodynamic simulations. This suggests that simulations predict a higher fraction of baryons residing in the IGM than analytic models.

\subsection{Mean Dispersion Measure}
\label{sec:mean_dm}
\begin{figure}
    \centering
    \includegraphics[width=\linewidth]{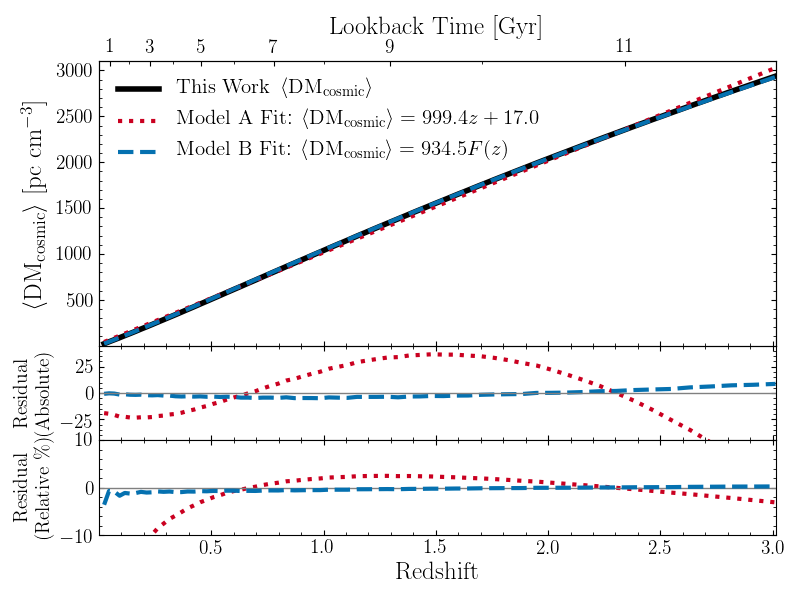}
    \caption{\emph{Top:} The solid black line is the $\left\langle \mathrm{DM_{cosmic}} \right\rangle$ at each redshift. The red and blue lines are the Model A (linear model) and Model B (non-linear model) fits the mean respectively. \emph{Middle:} The residuals of $\left\langle \mathrm{DM_{cosmic}} \right\rangle$ minus the model fits. The colours here are the same as the top panel. \emph{Bottom:} The relative residuals of the mean minus the model fits. The colours here are the same as the other panels.}
    \label{fig:mean_DM_fits}
\end{figure}

We calculated the mean $\left\langle \mathrm{DM_\mathrm{cosmo}} \right\rangle$ at each of the 66 redshift samples using the relation
\begin{equation}
    \langle \mathrm{DM_{cosmic}} \rangle = \sum_{i=0}^{N_\mathrm{bins}} \mathrm{DM}_i~P(\mathrm{DM}_i | z)~ \Delta \mathrm{DM}_i  \,,
    \label{eq:DM_Mean}
\end{equation}
where $\mathrm{DM}_i$ is a bin value, $P(\mathrm{DM}_i | z)$ is the probability of a line-of-sight with $\mathrm{DM}_i$ at redshift $z$ and $\Delta \mathrm{DM}_i$ is the width of the $\mathrm{DM}_i$ bin.

In the top \Cref{fig:mean_DM_fits} we show the measured $\left\langle \mathrm{DM_\mathrm{cosmic}} \right\rangle$ for each redshift. To fit this data we have considered both a linear and non-linear model for the $\mathrm{DM} - z$ relation.

We term the linear relation between $\left\langle \mathrm{DM_{cosmic}} \right\rangle$ and redshift as Model A where, 
\begin{equation}
    \left\langle \mathrm{DM_{cosmic}} \right\rangle = az + b  \,.
\end{equation}

This model has been used previously in the literature to provide a simple a 'rule-of-thumb' for converting $\mathrm{DM_{Obs}}$ into an estimated redshift \citep[e.g.][]{Zhang2018,Petroff2019}. 
We have fit this model with both a zero and non-zero intercept and find that $b \neq 0$ is preferred. We note however that most of the literature uses Model A with $b=0$ .

Model B is a non-linear parameterisation of the $\mathrm{DM} - z$ relation to include the cosmology of the Universe
\begin{equation}
    \left\langle \mathrm{DM_{cosmic}} \right\rangle = \alpha F(z)\,,
\end{equation}
here the parameter $F(z)$ is given by the following
\begin{equation}
    F(z) = \int_0^z\frac{1 + z}{\sqrt{(1+z)\Omega_m + \Omega_\Lambda }} \mathrm{d}z \,.
\end{equation}

Model B is based on \Cref{eq:DM_IGM} which relates $\left\langle \mathrm{DM_{cosmic}} \right\rangle$ to $z$. At redshift $z=0$, $F(z) = 1$. This means we are able to measure the redshift $z=0$ value of $\alpha = \frac{3 c \mathrm{H}_0 \Omega_b}{8 \pi G m_p} f_\mathrm{IGM}(z=0) \chi(z=0)$. 

We calculated a least squares fit for both models as a function of redshift $z$. 
In the top panel of \Cref{fig:mean_DM_fits} we present the best fitting parameters for both Model A and Model B. 
In the lower two panels of \Cref{fig:mean_DM_fits} we have also plotted the absolute and relative residuals to the best fitting model. 
We find that the best fit to Model A has parameters $a = 999 \pm 4$ and $b = 17 \pm 6$, with a reduced $\chi^2=809.3$ with 64 degrees of freedom and a mean relative residual of 14.6\% in the redshift range $z = 0 - 3$. The best fit to Model A with $b = 0$ has parameters $a = 1009 \pm 3$ with a reduced $\chi^2=899.2$ with 65 degrees of freedom. We find that the Model A fit with $b\neq0$ is preferred. 

The best fitting Model B has parameters $\alpha = 934.5 \pm 0.3$, with a reduced $\chi^2 = 13.2$ with 65 degrees of freedom and a mean relative residual of 0.732\%. 

We find that Model B is a better fit to the $\left<\mathrm{DM_{cosmic}} \right>$ particularly at low redshifts ($z < 0.5$). We would urge caution when using a linear model for estimating FRBs redshifts with low DM values. Using $\alpha = 934.5$ we find $f_\mathrm{IGM}(z=0) \chi(z=0) = 0.85$.

\subsection{Standard Deviation} \label{sec:sigma_dm}
We have use two different (but related) metrics to measure the scatter around $\left\langle \mathrm{DM_{cosmic}} \right\rangle$ at a given redshift. These metrics are:

\begin{enumerate}
    \item $\sigma_\mathrm{Var}$: the variance of the PDF, and
    \item $\sigma_\mathrm{CI}$: the 1 $\sigma$ (68\%) confidence interval.
\end{enumerate}

We have used two metrics because they both quantify the spread around the mean but in different ways. $\sigma_\mathrm{Var}$ quantifies the expected deviation from the mean, whereas $\sigma_\mathrm{CI}$ describes the width. These $\sigma_\mathrm{Var}$, $\sigma_\mathrm{CI}$ are defined as,

\begin{equation}
    \sigma_\mathrm{Var}^2(z) = \sum_{i=0}^{N_\mathrm{bins}} (\mathrm{DM}_i - \left\langle\mathrm{DM_{cosmic}}\right\rangle)^2 P(\mathrm{DM}_i|z)\Delta \mathrm{DM}_i \,,
    \label{eq:sigma_var}
\end{equation}

\begin{equation}
    \sigma_\mathrm{CI}(z) = \frac{\mathrm{DM}_{84}(z) - \mathrm{DM}_{16}(z)}{2}
    \label{eq:sigma_ci} \,.
\end{equation}

In \Cref{eq:sigma_var},  $\mathrm{DM}_i$ is a bin value, $\left\langle\mathrm{DM_{cosmic}}\right\rangle$ is mean of the PDF at redshift $z$,  $P(\mathrm{DM}_i | z)$ is the probability of a line-of-sight with $\mathrm{DM}_i$ at redshift $z$ and $\Delta \mathrm{DM}_i$ is the width of the bin. In \Cref{eq:sigma_ci} $\mathrm{DM}_{84}(z)$ and $\mathrm{DM}_{16}(z)$ are the upper and lower limits of the 68\% confidence interval respectively.

When the $\left\langle\mathrm{DM_{cosmic}}\right\rangle$ PDF is Gaussian, both $\sigma_\mathrm{Var}$ and $\sigma_\mathrm{CI}$ are identical. However for distributions that are skewed, $\sigma_\mathrm{Var} > \sigma_\mathrm{CI}$. The larger the difference between $\sigma_\mathrm{Var}$ and $\sigma_\mathrm{CI}$, the more skewed the distribution.

\begin{figure}
    \centering
    \includegraphics[width=\linewidth]{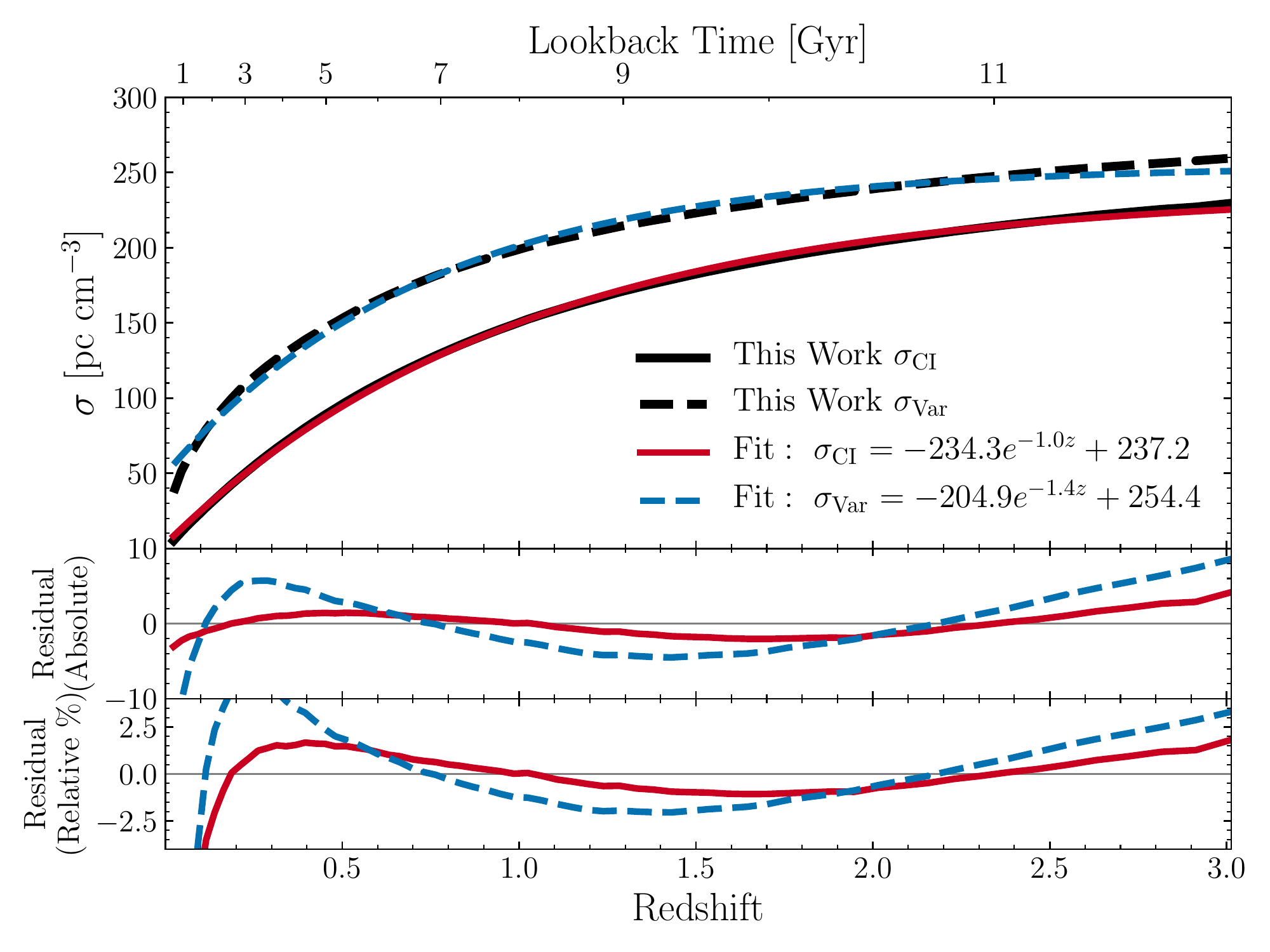}
    \caption{\emph{Top:} The solid and dashed black lines $\sigma_\mathrm{CI}$ and $\sigma_\mathrm{Var}$ at each redshift respectively. The red and blue lines are the best fitting exponential models for $\sigma_\mathrm{CI}$ (solid) and $\sigma_\mathrm{Var}$ (dashed) respectively. \emph{Middle:} The residuals of $\sigma_\mathrm{CI}$ and $\sigma_\mathrm{Var}$ minus the model fits. The line styles here are the same as the top panel. \emph{Bottom:} The relative residuals of $\sigma_\mathrm{CI}$ and $\sigma_\mathrm{Var}$ minus the model fits. The line styles here are the same as the other panels.}
    \label{fig:sigma_fits_v_redshift}
\end{figure}

In the top panel of \Cref{fig:sigma_fits_v_redshift} we show $\sigma_\mathrm{CI}$ and $\sigma_\mathrm{Var}$ for each redshift. 

We calculated a least squares fit for both $\sigma_\mathrm{Var}$ and $\sigma_\mathrm{CI}$ assuming an exponential form  with redshift $z$ as
\begin{equation}
    \sigma = Ae^{Bz} + C \,.
    \label{eq:expon}
\end{equation}

We plot the best fitting exponential models in \Cref{fig:sigma_fits_v_redshift}. In the lower two panels of \Cref{fig:sigma_fits_v_redshift} we have also plotted the absolute and relative residuals.
We find that the best exponential fit to
$\sigma_\mathrm{Var}$ has parameters $A=-205\pm2$, $B=-1.35\pm0.04$ and $C=254\pm2$ with a reduced $\chi^2=21.4$ with 63 degrees of freedom and a mean relative residual of 7.12\% in the redshift range $z = 0 - 3$.

Similarly the best fitting exponential to 
$\sigma_\mathrm{CI}$ has the parameters $A=-234.3\pm0.7$, $B = -0.991 \pm 0.009$ and $C = 237.2 \pm 0.8$ with a reduced $\chi^2=2.18$ with 63 degrees of freedom and a mean relative residual of 7.9\% in the redshift range $z = 0 - 3$.

We find that $\sigma_\mathrm{Var}$ is significantly larger than $\sigma_\mathrm{CI}$ for all redshifts indicating that the $\mathrm{DM_{cosmic}}$ PDF are non-Gaussian. 

\subsection{PDF Non-Gaussianity}
\begin{figure}
    \centering
    \includegraphics[width=\linewidth]{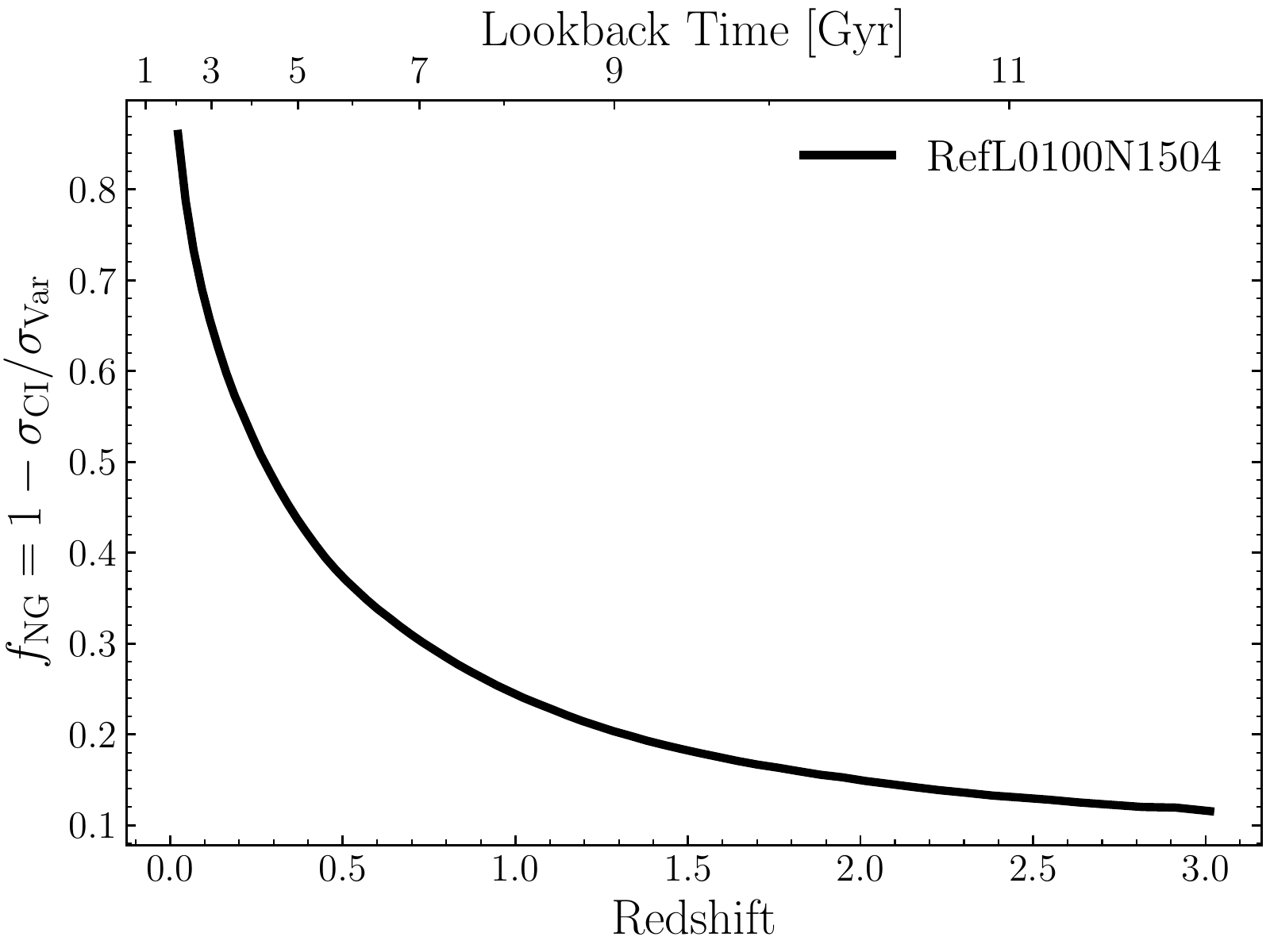}
    \caption{Non-Gaussainity of the PDFs, $f_\mathrm{NG}$, as a function of redshift, $z$. A value of $f_\mathrm{NG}$  of 0.0 indicates that the $\left\langle \mathrm{DM_{cosmic}}\right\rangle$ PDF at redshift $z$ is Gaussian. Ratios other than 0.0 indicate that the PDF is skewed from normality. The Gaussianity of the $\left\langle \mathrm{DM_{cosmic}}\right\rangle$ PDFs increases with redshift.}
    \label{fig:f_NG_v_redshift}
\end{figure}

We introduce the quantity $f_\mathrm{NG}$ as a measure of the ``non-Gaussianity" of the PDFs at any given redshift as described by
\begin{equation}
    f_\mathrm{NG} = 1 - \frac{\sigma_\mathrm{CI}}{\sigma_\mathrm{Var}} \,.
    \label{eq:f_NG}
\end{equation}

A value of $f_\mathrm{NG}$ that is close to zero indicates the PDF is close to Gaussian, whereas a value close to unity indicates extremely strong non-Gaussianity.  

In \Cref{fig:f_NG_v_redshift} we plot $f_\mathrm{NG}$ for each redshift. Over the redshift range $0 < z < 3$, we find that the Gaussianity of the PDFs increases markedly with redshift. We note that even at redshift $z = 3$, $f_\mathrm{NG}\approx 0.1$. This indicates that even at high redshifts ($z \sim 3$), the shape of the $\mathrm{DM_{cosmic}}$ is still significantly non-Gaussian. This increase in Gaussianity with redshift can be well interpreted in terms of the FRB path length and the amount of substructure up to that distance.

The shorter path length of low-redshift ($z < 0.5)$ FRBs increases the probability that the line-of-sight will not intersect with any galaxy halos or IGM filaments. The few high DM structures that intersect the sight lines of these low-redshift FRBs cause the PDF to become skewed to reflect the log-normal matter distribution.

At higher-redshifts ($z > 0.5$), the FRB sight lines are likely intersect with more high DM structures, causing the PDF to become more Gaussian with increasing redshift.

Additionally, the size of the simulation box contributes to the shape of the PDF distribution because it physically determines the maximum size of structures. A small, 25 cMpc box is physically unable to contain the low-density voids, or extremely rare high-density clusters because they would extend outside the box. Using larger boxs alleviates this problem because they are able to contain larger substructure. See \Cref{fig:conv_test_fng_v_redshift} for a comparison of $f_\mathrm{NG}$ with different simulation box sizes.

Another factor to consider is that different parameterisations of galaxy feedback mechanisms (i.e AGN and star formation) will lead to different growths of substructure. For example: increases in AGN feedback would likely eject more baryons into the IGM and decrease baryons in galaxies. The reduced amount of baryons in galaxies could cause the shape of the $\mathrm{DM_{cosmic}}$ to become more Gaussian at low redshifts ($z < 0.5$).

In the future as we move towards a situation of hundreds (or thousands) of localised FRBs, it may become possible to constrain the feedback mechanisms involved in galaxy evolution. The ability of future FRB host galaxy surveys to measure galaxy feedback processes will be explored more in Batten et al. \emph{in prep.}

\subsection{Comparison to Localised Fast Radio Bursts}
\begin{table*}
    \centering
    \begin{tabular}{cccccccl}
        \hline
         Name & R.A. (J2000) & Dec. (J2000) & $z_\mathrm{host}$ & $\mathrm{DM_{Obs}}$ & $\mathrm{DM_{MW}}$ & Repeater & \\
         & (Deg) & (Deg) & & $\left[\mathrm{pc\ cm^{-3}}\right]$ & $\left[\mathrm{pc\ cm^{-3}}\right]$ &  & \\
         \hline
         \hline
         FRB 180916 & 01:58:00.28 & $+$65:42:53.0 & 0.034 & 348.8 & 199 & Yes &\citet{Marcote2020} \\ 
         FRB 190608 & 22:16:04.90 & $-$07:53:55.8 & 0.118 & 339.5 & 37.2 & No &\citet{Macquart2020} \\
         FRB 200430 & 15:18:49.52 & $+$12:22:35.8 & 0.160 & 380.0 & 27.2 & No &\citet{Heintz2020}\\ 
         FRB 121102 & 05:31:58.70 & $+$33:08:52.7 & 0.193 & 557.0 & 188 & Yes &\citet{Tendulkar2017} \\
         FRB 191001 & 21:33:24.44 & $-$54:44:54.7 & 0.2340 & 507.9 & 44.2 & No & \citet{Bhandari2020}\\
         FRB 190714 & 12:15:55.09 & $-$13:01:16.0 & 0.2365 & 504.1 & 38.5 & No & \citet{Heintz2020}\\
         FRB 190102 & 21:29:39.72 & $-$79:28:32.2 & 0.291 & 364.5 & 57.3 &No &\citet{Macquart2020} \\
         FRB 180924 & 21:44:25.25 & $-$40:54:00.8 & 0.321 & 361.4 & 40.5 &No &\citet{Bannister2019} \\
         FRB 190611 & 21:22:58.71 & $-$79:23:49.6 & 0.378  & 321.4 & 57.8 &No &\citet{Macquart2020} \\
         FRB 181112 & 21:49:23.68 & $-$52:58:15.4 & 0.476 & 589.3 & 40.2 &No &\citet{Prochaska2019b} \\
         FRB 190711 & 21:57:40.63 & $-$80:21:29.3 & 0.522  & 593.1 & 56.5 & Yes &\citet{Macquart2020} \\
         FRB 190523 & 13:48:15.43 & $+$72:28:14.4 & 0.660  & 760.8 & 37.0 & No &\citet{Ravi2019} \\
         \hline
    \end{tabular}
    \caption{All FRBs for which host galaxies have been identified as of November 2020. From left to right the columns are: the name of the FRB, right ascension of the FRB in J2000 coordinates, declination of the FRB in J2000 coordinates the redshift of the FRB host galaxy ($z_\mathrm{host}$), the observed dispersion measure ($\mathrm{DM_{Obs}}$), the estimated Milky Way contribution calculated from the NE2001 \citep{NE2001} Galactic electron density model ($\mathrm{DM_{MW}}$), repeating status (Yes/No), and the reference to the FRB host galaxy measurement. The FRBs are ordered in increasing redshift. We have plotted these FRBs against the \EAGLE\ $\mathrm{DM_{cosmic}}$ relation in \Cref{fig:localised_frbs}.}
    \label{tab:localised_frbs}
\end{table*}

\begin{figure}
    \centering
    \includegraphics[width=\linewidth]{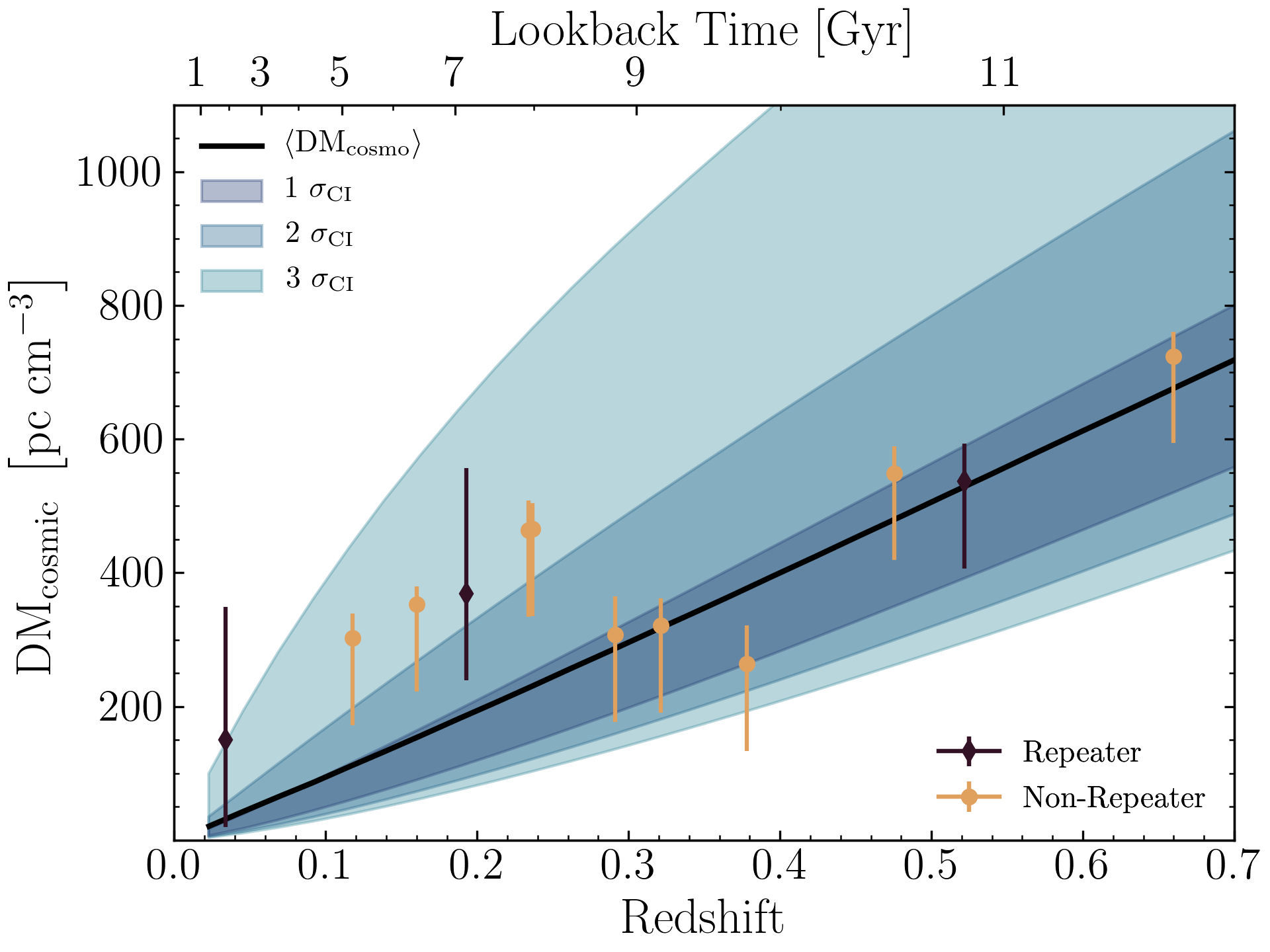}
    \caption{The solid black line is the mean $\mathrm{DM} - z$ relation found in this work shown in \Cref{fig:DMz_Hist}. The blue contours indicate the 1, 2 and 3 $\sigma_\mathrm{CI}$ confidence intervals. The data points show the DM versus redshift, $z$, for FRBs which have been associated with host galaxies (listed in \Cref{tab:localised_frbs}). For each FRB, the symbols show the DM excess ($\mathrm{DM_{obs}}$ - DM$_{\mathrm{MW}}$), the tip of the upward error bar shows the observed DM (DM$_{\mathrm{obs}}$). The Milky Way DM is estimated using the NE2001 galactic electron density model \citep{Cordes2019}. The tip of downward error bar indicates an additional contribution of 130 pc cm$^{-3}$ (composed of 100 pc cm$^{-3}$ for an estimated host galaxy and 30 pc cm$^{-3}$ for the contribution due to the Milky Way halo from \citealt{Prochaska2019a}; which is not modelled in NE2001). Diamonds (black) indicate known repeating FRBs, while filled circles (yellow) are FRBs that have not yet been seen to repeat.}
    \label{fig:localised_frbs}
\end{figure}

In \Cref{fig:localised_frbs} we plot $\left\langle \mathrm{DM_{cosmic}} \right\rangle$ from this work and the 1, 2 and 3 $\sigma_\mathrm{CI}$ confidence intervals. We clarify here that the $3 \sigma_\mathrm{CI}$ region is not simply $3 \times \sigma_\mathrm{CI}$, but the region that encompasses 99.7\% of the $\left\langle \mathrm{DM_{cosmic}} \right\rangle$ PDF.

We have also plotted the redshifts of the FRBs with localised host galaxies (see \Cref{tab:localised_frbs}). The tips of the error bar tops indicates $\mathrm{DM_{Obs}}$. The symbols here represent the DM excess ($\mathrm{DM_{Obs}} - \mathrm{DM_{MW}}$) with diamonds (black) and filled circles (yellow) indicating repeating and not observed to repeat FRBs respectively. We remind the reader that the DM excess of an FRB is not the same as $\mathrm{DM_{cosmic}}$. There are additional contributions to $\mathrm{DM_{Obs}}$ that are not model in the NE2001 electron density model. The two additional contributions that we consider are the halo of the Milky Way and the ionised gas inside of the FRBs host galaxy. For the Milky Way halo, we used the estimated 30 pc cm$^{-3}$ from \citet{Prochaska2019a}. We have excluded FRB~190614 \citep{Law2020} because the localisation identifies two possible host galaxies at redshift $z \sim 0.6$.  

The contribution due to ionised gas inside the host galaxy is unique to each FRB. We adopt a rather conservative estimate that each of the localised FRBs has a host galaxy contribution of 100 pc cm$^{-3}$. We expect that 100 pc cm$^{-3}$ is close to the upper limit for most FRBs based on the Milky Way DM contributions. This is assuming that there is not a large contribution to $\mathrm{DM_{Obs}}$ from the local environment surrounding the source of the FRB within the galaxy. 

The bottom of the error bars indicate an additional 130 pc cm$^{-3}$ of subtracted DM. This additional contribution accounts for contribution of ISM contribution 

We expect that for most FRBs, the DM excess falls above our mean $\mathrm{DM} - z$ relation when not accounting for the host galaxy contribution.

\Cref{fig:localised_frbs} shows that the $\mathrm{DM} - z$ relation obtained using the \EAGLE\ simulations broadly agrees with the observed FRBs host galaxies. We note that the number of localised FRBs is still small and a larger number is required, particularly at higher redshifts ($z \sim 2 - 3$). 

Of particular note that even with the significant amount of scatter we measure around $\left\langle \mathrm{DM_{cosmo}} \right\rangle$, the six FRBs with the lowest redshifts (FRB~180916, FRB~190608, FRB~200430, FRB~121102, FRB~191001 and FRB~190714) all reside in the $2-3\ \sigma_\mathrm{CI}$ regions, whereas the remaining high redshift FRBs more closely follow $\left\langle \mathrm{DM_{cosmic}} \right\rangle$.

There are a few reasons why an FRB would have a significantly larger $\mathrm{DM_{cosmic}}$ than is expected given its redshift. (i) the DM contribution from host galaxy and/or source environment around the FRB is larger than estimated, (ii) the FRB intersects with an unseen intervening galaxy halo, (iii) the FRB intersects with an over-dense filament of the IGM.

\citet{Simha2020} showed that FRB~190608 traverses through an over-dense filament of the IGM, which is why it has a DM excess much larger than the $\left\langle \mathrm{DM_{cosmic}} \right\rangle$ at redshift $z = 0.378$. 

We predict that if the IGM reconstruction similar to \citet{Simha2020} is performed on FRB~200430, FRB~191001 and FRB~190714 we expect that these will also indicate they intersect through an over-dense filament. We have not included the repeaters FRB~180916 and FRB~121102 as it is currently unclear if repeating FRBs and single burst FRBs have the same progenitors. A difference in progenitors could lead to significantly different source environment DM contributions.

On the other hand, it is unlikely that of the 12 localised FRB host galaxies, one quarter of them would intersect IGM filaments, particularly because they all originate from different locations on the sky. The large DM excess of the FRBs could instead be explained by significant host galaxy or source environment. 

\section{Summary and Conclusions}\label{sec:Conclusions}
In this paper we measured the DM for over 1 billion ($1504^2$) sight lines through the \EAGLE\ cosmological, hydrodynamic simulations. We used these sight lines to calculate the $\mathrm{DM} - z$ relation and the scatter around it for FRBs between redshifts $0 < z < 3$. We then compared our model with the observations of FRB host galaxy redshifts. We summarise our results as follows. 
\begin{itemize}
    \item We have fit the mean $\mathrm{DM}-z$ relation using both a linear relation (used often through the literature) and a non-linear relation that accounts for the cosmology of the Universe. We find that between redshifts $0 < z < 3$, the mean $\mathrm{DM} - z$ relations is best fit by the non-linear relation,
    \begin{equation}
        \left\langle \mathrm{DM_{cosmic}}\right\rangle = \left( 934.5 \pm 0.3 ~\mathrm{pc~cm^{-3}} \right) \int_0^z\frac{1 + z}{\sqrt{(1+z)\Omega_m + \Omega_\Lambda }} \mathrm{d}z \,.
    \end{equation}
    with $\Omega_m$ and $\Omega_\Lambda$ being 0.307 and 0.693 respectively. This fit has a mean residual of approximately 0.7\% in the redshift range $0 < z < 3$. 
    \item We find significant asymmetric scatter around the mean $\mathrm{DM} - z$ relation that increases exponentially with redshift. Due to the asymmetry in the $\mathrm{DM_{cosmic}}$ PDFs, we have used two different metrics to quantify the scatter around the mean $\sigma_\mathrm{Var}$ (the standard deviation) and $\sigma_\mathrm{CI}$ (the width of the 68\% confidence interval). The best fitting relations (exponential with $z$) for $\sigma_\mathrm{Var}$ and $\sigma_\mathrm{CI}$ are: 
    \begin{align}
        &\sigma_\mathrm{Var} = (-205\pm 2) \exp[(-1.35\pm 0.04)z] + (254 \pm 2)   \\
        &\sigma_\mathrm{CI} = (-237.3\pm 0.7) \exp[(-0.991\pm 0.009)z] + (237.2\pm 0.8)
    \end{align}

    These fits have a mean residual of approximately 7-8\% in the redshift range $0 < z < 3$. 
    \item We find that the $\mathrm{DM_{cosmic}}$ PDFs are strongly asymmetric  at low-redshifts ($z < 0.5$) and become more Gaussian as the redshift approaches $z \sim 3$.
    
    This is explained by the increasing path length of high redshift FRBs intersecting with many more high DM structures, causing the PDF to become more Gaussian. 
    However, the box size of the simulation is also an important factor. The smaller the simulation, the faster the PDFs will become Gaussian. This is because the box size physically constrains the maximum size of structure in the Universe. The simulations with box sizes of 25 cMpc can not contain clusters and voids of that scale or larger. Large simulations of the order 100 cMpc is required to measure the effects of the log-normal matter density profile on the shape of the $\mathrm{DM_{cosmic}}$ PDFs. 
    
    \item Even with the large amount of scatter, we find that the six FRBs with the closest host galaxies in redshift all have $\mathrm{DM_{cosmic}}$ values that places them in the $2-3\sigma$ confidence interval above the mean $\left\langle\mathrm{DM_{cosmic}}\right\rangle$. These FRBs include: FRB~180916, FRB~190608, FRB~200430, FRB~121102, FRB~191001 and FRB~190714. We predict that when an IGM reconstruction similar to the work of \citet{Simha2020} is performed on FRB~200430, FRB~191001 and FRB~190714, it will indicate that these FRBs also intersect a filament of the IGM. Alternatively, if these FRBs are not found to have intersected with an IGM filament, it would indicate a significant host galaxy or local environment DM contribution.
    \item We have made the $\mathrm{DM}-z$ relation published in this work publicly available through inclusion in the open source FRB redshift estimation package \textsc{fruitbat} \citep{Batten2019} which is available via pip and \mbox{\url{https://github.com/abatten/fruitbat}}.
\end{itemize}

\section*{Acknowledgements}
AJB would like to thank Chris Blake for the many helpful discussions and feedback he has given to the results presented in this work. We acknowledge the Wurundjeri People, the traditional owners of the land upon on which The Swinburne University of Technology (Hawthorn) is located. This research was supported by the Australian Research Council Centre of Excellence for All Sky Astrophysics in 3Dimensions (ASTRO 3D), through project number CE170100013. This work was performed on the OzSTAR national facility at Swinburne University of Technology. The OzSTAR program receives funding in part from the Astronomy National Collaborative Research Infrastructure Strategy (NCRIS) allocation provided by the Australian Government. This research has made use of NASA's Astrophysics Data System. This research made use of \texttt{matplotlib} \citep{Hunter:2007}, \texttt{SciPy} \citep{Virtanen_2020}, \texttt{IPython} \citep{PER-GRA:2007}, \texttt{Astropy} \citep{2013A&A...558A..33A, 2018AJ....156..123A}, \texttt{NumPy} \citep{van2011numpy, harris2020array}, \texttt{Pandas} \citep{McKinney_2010, McKinney_2011} and \texttt{CMasher} for scientific colour maps \citep{CMasher}. This paper made use of WebPlotDigitizer (http://arohatgi.info/WebPlotDigitizer/) by Ankit Rohatgi.



\bibliographystyle{mnras}
\bibliography{references} 

\begin{thebibliography}{}
\makeatletter
\relax
\def\mn@urlcharsother{\let\do\@makeother \do\$\do\&\do\#\do\^\do\_\do\%\do\~}
\def\mn@doi{\begingroup\mn@urlcharsother \@ifnextchar [ {\mn@doi@}
  {\mn@doi@[]}}
\def\mn@doi@[#1]#2{\def\@tempa{#1}\ifx\@tempa\@empty \href
  {http://dx.doi.org/#2} {doi:#2}\else \href {http://dx.doi.org/#2} {#1}\fi
  \endgroup}
\def\mn@eprint#1#2{\mn@eprint@#1:#2::\@nil}
\def\mn@eprint@arXiv#1{\href {http://arxiv.org/abs/#1} {{\tt arXiv:#1}}}
\def\mn@eprint@dblp#1{\href {http://dblp.uni-trier.de/rec/bibtex/#1.xml}
  {dblp:#1}}
\def\mn@eprint@#1:#2:#3:#4\@nil{\def\@tempa {#1}\def\@tempb {#2}\def\@tempc
  {#3}\ifx \@tempc \@empty \let \@tempc \@tempb \let \@tempb \@tempa \fi \ifx
  \@tempb \@empty \def\@tempb {arXiv}\fi \@ifundefined
  {mn@eprint@\@tempb}{\@tempb:\@tempc}{\expandafter \expandafter \csname
  mn@eprint@\@tempb\endcsname \expandafter{\@tempc}}}

\bibitem[\protect\citeauthoryear{{Astropy Collaboration} et~al.,}{{Astropy
  Collaboration} et~al.}{2013}]{2013A&A...558A..33A}
{Astropy Collaboration} et~al., 2013, \mn@doi [\aap]
  {10.1051/0004-6361/201322068}, \href
  {http://adsabs.harvard.edu/abs/2013A\%26A...558A..33A} {558, A33}

\bibitem[\protect\citeauthoryear{{Astropy Collaboration} et~al.,}{{Astropy
  Collaboration} et~al.}{2018}]{2018AJ....156..123A}
{Astropy Collaboration} et~al., 2018, \mn@doi [\aj] {10.3847/1538-3881/aabc4f},
  \href {http://adsabs.harvard.edu/abs/2018AJ....156..123A} {156, 123}

\bibitem[\protect\citeauthoryear{{Bannister} et~al.,}{{Bannister}
  et~al.}{2019}]{Bannister2019}
{Bannister} K.~W.,  et~al., 2019, \mn@doi [Science] {10.1126/science.aaw5903},
  \href {https://ui.adsabs.harvard.edu/abs/2019Sci...365..565B} {365, 565}

\bibitem[\protect\citeauthoryear{{Batten}}{{Batten}}{2019}]{Batten2019}
{Batten} A.~J.,  2019, \mn@doi [The Journal of Open Source Software]
  {10.21105/joss.01399}, \href
  {https://ui.adsabs.harvard.edu/abs/2019JOSS....4.1399B} {4, 1399}

\bibitem[\protect\citeauthoryear{{Bhandari} et~al.,}{{Bhandari}
  et~al.}{2018}]{Bhandari2018}
{Bhandari} S.,  et~al., 2018, \mn@doi [\mnras] {10.1093/mnras/stx3074}, \href
  {https://ui.adsabs.harvard.edu/abs/2018MNRAS.475.1427B} {475, 1427}

\bibitem[\protect\citeauthoryear{{Bhandari} et~al.,}{{Bhandari}
  et~al.}{2020}]{Bhandari2020}
{Bhandari} S.,  et~al., 2020, \mn@doi [\apjl] {10.3847/2041-8213/abb462}, \href
  {https://ui.adsabs.harvard.edu/abs/2020ApJ...901L..20B} {901, L20}

\bibitem[\protect\citeauthoryear{{Blaizot}, {Wadadekar}, {Guiderdoni},
  {Colombi}, {Bertin}, {Bouchet}, {Devriendt}  \& {Hatton}}{{Blaizot}
  et~al.}{2005}]{Blaizot2005}
{Blaizot} J.,  {Wadadekar} Y.,  {Guiderdoni} B.,  {Colombi} S.~T.,  {Bertin}
  E.,  {Bouchet} F.~R.,  {Devriendt} J. E.~G.,   {Hatton} S.,  2005, \mn@doi
  [\mnras] {10.1111/j.1365-2966.2005.09019.x}, \href
  {https://ui.adsabs.harvard.edu/abs/2005MNRAS.360..159B} {360, 159}

\bibitem[\protect\citeauthoryear{{Booth} \& {Schaye}}{{Booth} \&
  {Schaye}}{2009}]{Booth2009}
{Booth} C.~M.,  {Schaye} J.,  2009, \mn@doi [\mnras]
  {10.1111/j.1365-2966.2009.15043.x}, \href
  {https://ui.adsabs.harvard.edu/abs/2009MNRAS.398...53B} {398, 53}

\bibitem[\protect\citeauthoryear{{Bregman}}{{Bregman}}{2007}]{Bregman2007}
{Bregman} J.~N.,  2007, \mn@doi [\araa]
  {10.1146/annurev.astro.45.051806.110619}, \href
  {https://ui.adsabs.harvard.edu/abs/2007ARA&A..45..221B} {45, 221}

\bibitem[\protect\citeauthoryear{{Cen} \& {Ostriker}}{{Cen} \&
  {Ostriker}}{1999}]{Cen1999}
{Cen} R.,  {Ostriker} J.~P.,  1999, \mn@doi [\apj] {10.1086/306949}, \href
  {https://ui.adsabs.harvard.edu/abs/1999ApJ...514....1C} {514, 1}

\bibitem[\protect\citeauthoryear{{Chabrier}}{{Chabrier}}{2003}]{Chabrier2003}
{Chabrier} G.,  2003, \mn@doi [\pasp] {10.1086/376392}, \href
  {https://ui.adsabs.harvard.edu/abs/2003PASP..115..763C} {115, 763}

\bibitem[\protect\citeauthoryear{Cordes \& Chatterjee}{Cordes \&
  Chatterjee}{2019}]{Cordes2019}
Cordes J.~M.,  Chatterjee S.,  2019, \mn@doi [Annual Review of Astronomy and
  Astrophysics] {10.1146/annurev-astro-091918-104501}, 57, 417

\bibitem[\protect\citeauthoryear{{Cordes} \& {Lazio}}{{Cordes} \&
  {Lazio}}{2002}]{NE2001}
{Cordes} J.~M.,  {Lazio} T.~J.~W.,  2002, arXiv e-prints, \href
  {https://ui.adsabs.harvard.edu/abs/2002astro.ph..7156C} {pp
  astro--ph/0207156}

\bibitem[\protect\citeauthoryear{{Crain} et~al.,}{{Crain}
  et~al.}{2015}]{Crain2015}
{Crain} R.~A.,  et~al., 2015, \mn@doi [\mnras] {10.1093/mnras/stv725}, \href
  {https://ui.adsabs.harvard.edu/abs/2015MNRAS.450.1937C} {450, 1937}

\bibitem[\protect\citeauthoryear{Dalla~Vecchia \& Schaye}{Dalla~Vecchia \&
  Schaye}{2012}]{DallaVecchia2012}
Dalla~Vecchia C.,  Schaye J.,  2012, \mn@doi [Monthly Notices of the Royal
  Astronomical Society] {10.1111/j.1365-2966.2012.21704.x}, 426, 140

\bibitem[\protect\citeauthoryear{{Deng} \& {Zhang}}{{Deng} \&
  {Zhang}}{2014}]{Deng2014}
{Deng} W.,  {Zhang} B.,  2014, \mn@doi [\apjl] {10.1088/2041-8205/783/2/L35},
  \href {https://ui.adsabs.harvard.edu/abs/2014ApJ...783L..35D} {783, L35}

\bibitem[\protect\citeauthoryear{Dolag, Gaensler, Beck  \& Beck}{Dolag
  et~al.}{2015}]{Dolag2015}
Dolag K.,  Gaensler B.~M.,  Beck A.~M.,   Beck M.~C.,  2015, \mn@doi [\mnras]
  {10.1093/mnras/stv1190}, 451, 4277

\bibitem[\protect\citeauthoryear{{Ferland}, {Korista}, {Verner}, {Ferguson},
  {Kingdon}  \& {Verner}}{{Ferland} et~al.}{1998}]{Ferland1998}
{Ferland} G.~J.,  {Korista} K.~T.,  {Verner} D.~A.,  {Ferguson} J.~W.,
  {Kingdon} J.~B.,   {Verner} E.~M.,  1998, \mn@doi [\pasp] {10.1086/316190},
  \href {https://ui.adsabs.harvard.edu/abs/1998PASP..110..761F} {110, 761}

\bibitem[\protect\citeauthoryear{{Fosalba}, {Gazta{\~n}aga}, {Castand er}  \&
  {Manera}}{{Fosalba} et~al.}{2008}]{Fosalba2008}
{Fosalba} P.,  {Gazta{\~n}aga} E.,  {Castand er} F.~J.,   {Manera} M.,  2008,
  \mn@doi [\mnras] {10.1111/j.1365-2966.2008.13910.x}, \href
  {https://ui.adsabs.harvard.edu/abs/2008MNRAS.391..435F} {391, 435}

\bibitem[\protect\citeauthoryear{{Furlong} et~al.,}{{Furlong}
  et~al.}{2015}]{Furlong2015}
{Furlong} M.,  et~al., 2015, \mn@doi [\mnras] {10.1093/mnras/stv852}, \href
  {https://ui.adsabs.harvard.edu/abs/2015MNRAS.450.4486F} {450, 4486}

\bibitem[\protect\citeauthoryear{{Haardt} \& {Madau}}{{Haardt} \&
  {Madau}}{2001}]{HaardtMadau2001}
{Haardt} F.,  {Madau} P.,  2001, in {Neumann} D.~M.,  {Tran} J.~T.~V.,  eds,
  Clusters of Galaxies and the High Redshift Universe Observed in X-rays. p.~64
  (\mn@eprint {arXiv} {astro-ph/0106018})

\bibitem[\protect\citeauthoryear{Harris et~al.,}{Harris
  et~al.}{2020}]{harris2020array}
Harris C.~R.,  et~al., 2020, \mn@doi [Nature] {10.1038/s41586-020-2649-2}, 585,
  357

\bibitem[\protect\citeauthoryear{{Heintz} et~al.,}{{Heintz}
  et~al.}{2020}]{Heintz2020}
{Heintz} K.~E.,  et~al., 2020, arXiv e-prints, \href
  {https://ui.adsabs.harvard.edu/abs/2020arXiv200910747H} {p. arXiv:2009.10747}

\bibitem[\protect\citeauthoryear{Hunter}{Hunter}{2007}]{Hunter:2007}
Hunter J.~D.,  2007, Computing In Science \& Engineering, 9, 90

\bibitem[\protect\citeauthoryear{{Inoue}}{{Inoue}}{2004}]{Inoue2004}
{Inoue} S.,  2004, \mn@doi [\mnras] {10.1111/j.1365-2966.2004.07359.x}, \href
  {https://ui.adsabs.harvard.edu/abs/2004MNRAS.348..999I} {348, 999}

\bibitem[\protect\citeauthoryear{{Ioka}}{{Ioka}}{2003}]{Ioka2003}
{Ioka} K.,  2003, \mn@doi [\apjl] {10.1086/380598}, \href
  {https://ui.adsabs.harvard.edu/abs/2003ApJ...598L..79I} {598, L79}

\bibitem[\protect\citeauthoryear{{Jaroszynski}}{{Jaroszynski}}{2019}]{Jaroszynski2019}
{Jaroszynski} M.,  2019, \mn@doi [\mnras] {10.1093/mnras/sty3529}, \href
  {https://ui.adsabs.harvard.edu/abs/2019MNRAS.484.1637J} {484, 1637}

\bibitem[\protect\citeauthoryear{{Keane}}{{Keane}}{2019}]{Keane2019}
{Keane} E.~F.,  2019, \mn@doi [Nature Astronomy] {10.1038/s41550-018-0603-0},
  \href {https://ui.adsabs.harvard.edu/abs/2018NatAs...2..865K} {2, 865}

\bibitem[\protect\citeauthoryear{{Kennicutt}}{{Kennicutt}}{1998}]{Kennicutt1998}
{Kennicutt} Robert~C. J.,  1998, \mn@doi [\apj] {10.1086/305588}, \href
  {https://ui.adsabs.harvard.edu/abs/1998ApJ...498..541K} {498, 541}

\bibitem[\protect\citeauthoryear{{Law} et~al.,}{{Law} et~al.}{2020}]{Law2020}
{Law} C.~J.,  et~al., 2020, \mn@doi [\apj] {10.3847/1538-4357/aba4ac}, \href
  {https://ui.adsabs.harvard.edu/abs/2020ApJ...899..161L} {899, 161}

\bibitem[\protect\citeauthoryear{{Lorimer}, {Bailes}, {McLaughlin}, {Narkevic}
  \& {Crawford}}{{Lorimer} et~al.}{2007}]{Lorimer2007}
{Lorimer} D.~R.,  {Bailes} M.,  {McLaughlin} M.~A.,  {Narkevic} D.~J.,
  {Crawford} F.,  2007, \mn@doi [Science] {10.1126/science.1147532}, \href
  {https://ui.adsabs.harvard.edu/abs/2007Sci...318..777L} {318, 777}

\bibitem[\protect\citeauthoryear{{Macquart} et~al.,}{{Macquart}
  et~al.}{2020}]{Macquart2020}
{Macquart} J.~P.,  et~al., 2020, \mn@doi [\nat] {10.1038/s41586-020-2300-2},
  \href {https://ui.adsabs.harvard.edu/abs/2020Natur.581..391M} {581, 391}

\bibitem[\protect\citeauthoryear{{Madau}, {Ferguson}, {Dickinson},
  {Giavalisco}, {Steidel}  \& {Fruchter}}{{Madau} et~al.}{1996}]{Madau1996}
{Madau} P.,  {Ferguson} H.~C.,  {Dickinson} M.~E.,  {Giavalisco} M.,  {Steidel}
  C.~C.,   {Fruchter} A.,  1996, \mn@doi [\mnras] {10.1093/mnras/283.4.1388},
  \href {https://ui.adsabs.harvard.edu/abs/1996MNRAS.283.1388M} {283, 1388}

\bibitem[\protect\citeauthoryear{{Marcote} et~al.,}{{Marcote}
  et~al.}{2017}]{Marcote2017}
{Marcote} B.,  et~al., 2017, \mn@doi [\apjl] {10.3847/2041-8213/834/2/L8},
  \href {https://ui.adsabs.harvard.edu/abs/2017ApJ...834L...8M} {834, L8}

\bibitem[\protect\citeauthoryear{{Marcote} et~al.,}{{Marcote}
  et~al.}{2020}]{Marcote2020}
{Marcote} B.,  et~al., 2020, \mn@doi [\nat] {10.1038/s41586-019-1866-z}, \href
  {https://ui.adsabs.harvard.edu/abs/2020Natur.577..190M} {577, 190}

\bibitem[\protect\citeauthoryear{McKinney}{McKinney}{2010}]{McKinney_2010}
McKinney W.,  2010, in Proceedings of the 9th Python in Science Conference. pp
  51--56

\bibitem[\protect\citeauthoryear{McKinney}{McKinney}{2011}]{McKinney_2011}
McKinney W.,  2011, Python for High Performance and Scientific Computing, 14

\bibitem[\protect\citeauthoryear{{McQuinn}}{{McQuinn}}{2014}]{McQuinn2014}
{McQuinn} M.,  2014, \mn@doi [\apjl] {10.1088/2041-8205/780/2/L33}, \href
  {https://ui.adsabs.harvard.edu/abs/2014ApJ...780L..33M} {780, L33}

\bibitem[\protect\citeauthoryear{P\'erez \& Granger}{P\'erez \&
  Granger}{2007}]{PER-GRA:2007}
P\'erez F.,  Granger B.~E.,  2007, \mn@doi [Computing in Science and
  Engineering] {10.1109/MCSE.2007.53}, 9, 21

\bibitem[\protect\citeauthoryear{{Petroff} et~al.,}{{Petroff}
  et~al.}{2016}]{Petroff2016}
{Petroff} E.,  et~al., 2016, \mn@doi [\pasa] {10.1017/pasa.2016.35}, \href
  {https://ui.adsabs.harvard.edu/abs/2016PASA...33...45P} {33, e045}

\bibitem[\protect\citeauthoryear{{Petroff}, {Hessels}  \& {Lorimer}}{{Petroff}
  et~al.}{2019}]{Petroff2019}
{Petroff} E.,  {Hessels} J.~W.~T.,   {Lorimer} D.~R.,  2019, \mn@doi [\aapr]
  {10.1007/s00159-019-0116-6}, \href
  {https://ui.adsabs.harvard.edu/abs/2019A&ARv..27....4P} {27, 4}

\bibitem[\protect\citeauthoryear{{Planck Collaboration} et~al.,}{{Planck
  Collaboration} et~al.}{2014}]{Planck2014}
{Planck Collaboration} et~al., 2014, \mn@doi [A\&A]
  {10.1051/0004-6361/201321529}, \href
  {https://ui.adsabs.harvard.edu/\#abs/2014A&A...571A...1P} {571, A1}

\bibitem[\protect\citeauthoryear{{Pol}, {Lam}, {McLaughlin}, {Lazio}  \&
  {Cordes}}{{Pol} et~al.}{2019}]{Pol2019}
{Pol} N.,  {Lam} M.~T.,  {McLaughlin} M.~A.,  {Lazio} T.~J.~W.,   {Cordes}
  J.~M.,  2019, \mn@doi [\apj] {10.3847/1538-4357/ab4c2f}, \href
  {https://ui.adsabs.harvard.edu/abs/2019ApJ...886..135P} {886, 135}

\bibitem[\protect\citeauthoryear{{Prochaska} \& {Zheng}}{{Prochaska} \&
  {Zheng}}{2019}]{Prochaska2019a}
{Prochaska} J.~X.,  {Zheng} Y.,  2019, \mn@doi [\mnras] {10.1093/mnras/stz261},
  \href {https://ui.adsabs.harvard.edu/abs/2019MNRAS.485..648P} {485, 648}

\bibitem[\protect\citeauthoryear{{Prochaska} et~al.,}{{Prochaska}
  et~al.}{2013}]{Prochaska2013}
{Prochaska} J.~X.,  et~al., 2013, \mn@doi [\apj] {10.1088/0004-637X/776/2/136},
  \href {https://ui.adsabs.harvard.edu/abs/2013ApJ...776..136P} {776, 136}

\bibitem[\protect\citeauthoryear{{Prochaska} et~al.,}{{Prochaska}
  et~al.}{2019}]{Prochaska2019b}
{Prochaska} J.~X.,  et~al., 2019, \mn@doi [Science] {10.1126/science.aay0073},
  \href {https://ui.adsabs.harvard.edu/abs/2019Sci...365.0073P} {365, aay0073}

\bibitem[\protect\citeauthoryear{{Rahmati}, {Pawlik}, {Rai{\v{c}}evi{\'c}}  \&
  {Schaye}}{{Rahmati} et~al.}{2013}]{Rahmati2013}
{Rahmati} A.,  {Pawlik} A.~H.,  {Rai{\v{c}}evi{\'c}} M.,   {Schaye} J.,  2013,
  \mn@doi [\mnras] {10.1093/mnras/stt066}, \href
  {https://ui.adsabs.harvard.edu/abs/2013MNRAS.430.2427R} {430, 2427}

\bibitem[\protect\citeauthoryear{{Rahmati}, {Schaye}, {Bower}, {Crain},
  {Furlong}, {Schaller}  \& {Theuns}}{{Rahmati} et~al.}{2015}]{Rahmati2015}
{Rahmati} A.,  {Schaye} J.,  {Bower} R.~G.,  {Crain} R.~A.,  {Furlong} M.,
  {Schaller} M.,   {Theuns} T.,  2015, \mn@doi [\mnras]
  {10.1093/mnras/stv1414}, \href
  {https://ui.adsabs.harvard.edu/abs/2015MNRAS.452.2034R} {452, 2034}

\bibitem[\protect\citeauthoryear{{Ravi} et~al.,}{{Ravi}
  et~al.}{2019}]{Ravi2019}
{Ravi} V.,  et~al., 2019, \mn@doi [\nat] {10.1038/s41586-019-1389-7}, \href
  {https://ui.adsabs.harvard.edu/abs/2019Natur.572..352R} {572, 352}

\bibitem[\protect\citeauthoryear{{Rudie} et~al.,}{{Rudie}
  et~al.}{2012}]{Rudie2012}
{Rudie} G.~C.,  et~al., 2012, \mn@doi [\apj] {10.1088/0004-637X/750/1/67},
  \href {https://ui.adsabs.harvard.edu/abs/2012ApJ...750...67R} {750, 67}

\bibitem[\protect\citeauthoryear{{Schaye} \& {Dalla Vecchia}}{{Schaye} \&
  {Dalla Vecchia}}{2008}]{Schaye2008}
{Schaye} J.,  {Dalla Vecchia} C.,  2008, \mn@doi [\mnras]
  {10.1111/j.1365-2966.2007.12639.x}, \href
  {https://ui.adsabs.harvard.edu/abs/2008MNRAS.383.1210S} {383, 1210}

\bibitem[\protect\citeauthoryear{Schaye et~al.,}{Schaye
  et~al.}{2015}]{Schaye2015}
Schaye J.,  et~al., 2015, \mn@doi [\mnras.] {10.1093/mnras/stu2058}, 446, 521

\bibitem[\protect\citeauthoryear{{Shull}, {Smith}  \& {Danforth}}{{Shull}
  et~al.}{2012}]{Shull2012}
{Shull} J.~M.,  {Smith} B.~D.,   {Danforth} C.~W.,  2012, \mn@doi [\apj]
  {10.1088/0004-637X/759/1/23}, \href
  {https://ui.adsabs.harvard.edu/abs/2012ApJ...759...23S} {759, 23}

\bibitem[\protect\citeauthoryear{{Simha} et~al.,}{{Simha}
  et~al.}{2020}]{Simha2020}
{Simha} S.,  et~al., 2020, \mn@doi [\apj] {10.3847/1538-4357/abafc3}, \href
  {https://ui.adsabs.harvard.edu/abs/2020ApJ...901..134S} {901, 134}

\bibitem[\protect\citeauthoryear{{Spitler} et~al.,}{{Spitler}
  et~al.}{2014}]{Spitler2014}
{Spitler} L.~G.,  et~al., 2014, \mn@doi [\apj] {10.1088/0004-637X/790/2/101},
  \href {https://ui.adsabs.harvard.edu/abs/2014ApJ...790..101S} {790, 101}

\bibitem[\protect\citeauthoryear{{Spitler} et~al.,}{{Spitler}
  et~al.}{2016}]{Spitler2016}
{Spitler} L.~G.,  et~al., 2016, \mn@doi [\nat] {10.1038/nature17168}, \href
  {https://ui.adsabs.harvard.edu/abs/2016Natur.531..202S} {531, 202}

\bibitem[\protect\citeauthoryear{{Springel}}{{Springel}}{2005}]{Springel2005}
{Springel} V.,  2005, \mn@doi [\mnras] {10.1111/j.1365-2966.2005.09655.x},
  \href {https://ui.adsabs.harvard.edu/abs/2005MNRAS.364.1105S} {364, 1105}

\bibitem[\protect\citeauthoryear{{Tendulkar} et~al.,}{{Tendulkar}
  et~al.}{2017}]{Tendulkar2017}
{Tendulkar} S.~P.,  et~al., 2017, \mn@doi [\apjl] {10.3847/2041-8213/834/2/L7},
  \href {https://ui.adsabs.harvard.edu/abs/2017ApJ...834L...7T} {834, L7}

\bibitem[\protect\citeauthoryear{{The CHIME/FRB Collaboration} et~al.,}{{The
  CHIME/FRB Collaboration} et~al.}{2019a}]{CHIME2019_Repeaters}
{The CHIME/FRB Collaboration} et~al., 2019a, arXiv e-prints, \href
  {https://ui.adsabs.harvard.edu/abs/2019arXiv190803507T} {p. arXiv:1908.03507}

\bibitem[\protect\citeauthoryear{{The CHIME/FRB Collaboration} et~al.,}{{The
  CHIME/FRB Collaboration} et~al.}{2019b}]{CHIME2019_second_repeater}
{The CHIME/FRB Collaboration} et~al., 2019b, \mn@doi [\nat]
  {10.1038/s41586-018-0864-x}, \href
  {https://ui.adsabs.harvard.edu/abs/2019Natur.566..235C} {566, 235}

\bibitem[\protect\citeauthoryear{Thornton et~al.,}{Thornton
  et~al.}{2013}]{Thornton2013}
Thornton D.,  et~al., 2013, \mn@doi [Science] {10.1126/science.1236789}, 341,
  53

\bibitem[\protect\citeauthoryear{Van Der~Walt, Colbert  \& Varoquaux}{Van
  Der~Walt et~al.}{2011}]{van2011numpy}
Van Der~Walt S.,  Colbert S.~C.,   Varoquaux G.,  2011, Computing in Science \&
  Engineering, 13, 22

\bibitem[\protect\citeauthoryear{{Virtanen} et~al.,}{{Virtanen}
  et~al.}{2020}]{Virtanen_2020}
{Virtanen} P.,  et~al., 2020, \mn@doi [Nature Methods]
  {https://doi.org/10.1038/s41592-019-0686-2}, \href {https://rdcu.be/b08Wh}
  {17, 261}

\bibitem[\protect\citeauthoryear{{Wiersma}, {Schaye}  \& {Smith}}{{Wiersma}
  et~al.}{2009a}]{Wiersma2009a}
{Wiersma} R. P.~C.,  {Schaye} J.,   {Smith} B.~D.,  2009a, \mn@doi [\mnras]
  {10.1111/j.1365-2966.2008.14191.x}, \href
  {https://ui.adsabs.harvard.edu/abs/2009MNRAS.393...99W} {393, 99}

\bibitem[\protect\citeauthoryear{{Wiersma}, {Schaye}, {Theuns}, {Dalla Vecchia}
   \& {Tornatore}}{{Wiersma} et~al.}{2009b}]{Wiersma2009b}
{Wiersma} R. P.~C.,  {Schaye} J.,  {Theuns} T.,  {Dalla Vecchia} C.,
  {Tornatore} L.,  2009b, \mn@doi [\mnras] {10.1111/j.1365-2966.2009.15331.x},
  \href {https://ui.adsabs.harvard.edu/abs/2009MNRAS.399..574W} {399, 574}

\bibitem[\protect\citeauthoryear{{Wijers}, {Schaye}, {Oppenheimer}, {Crain}  \&
  {Nicastro}}{{Wijers} et~al.}{2019}]{Wijers2019}
{Wijers} N.~A.,  {Schaye} J.,  {Oppenheimer} B.~D.,  {Crain} R.~A.,
  {Nicastro} F.,  2019, \mn@doi [\mnras] {10.1093/mnras/stz1762}, \href
  {https://ui.adsabs.harvard.edu/abs/2019MNRAS.488.2947W} {488, 2947}

\bibitem[\protect\citeauthoryear{{Yang}, {Luo}, {Li}  \& {Zhang}}{{Yang}
  et~al.}{2017}]{Yang2017}
{Yang} Y.-P.,  {Luo} R.,  {Li} Z.,   {Zhang} B.,  2017, \mn@doi [\apjl]
  {10.3847/2041-8213/aa6c2e}, \href
  {https://ui.adsabs.harvard.edu/abs/2017ApJ...839L..25Y} {839, L25}

\bibitem[\protect\citeauthoryear{{Yao}, {Manchester}  \& {Wang}}{{Yao}
  et~al.}{2017}]{YMW2016}
{Yao} J.~M.,  {Manchester} R.~N.,   {Wang} N.,  2017, \mn@doi [\apj]
  {10.3847/1538-4357/835/1/29}, \href
  {https://ui.adsabs.harvard.edu/abs/2017ApJ...835...29Y} {835, 29}

\bibitem[\protect\citeauthoryear{Zhang}{Zhang}{2018}]{Zhang2018}
Zhang B.,  2018, \mn@doi [\apj] {10.3847/2041-8213/aae8e3}, 867, L21

\bibitem[\protect\citeauthoryear{{van der Velden}}{{van der
  Velden}}{2020}]{CMasher}
{van der Velden} E.,  2020, \mn@doi [The Journal of Open Source Software]
  {10.21105/joss.02004}, \href
  {https://ui.adsabs.harvard.edu/abs/2020JOSS....5.2004V} {5, 2004}

\makeatother
\end{thebibliography}




\appendix
\section{Model Availability in \textsc{fruitbat}}
We have made the $\mathrm{DM}-z$ relation from this work available in the open source python package \textsc{fruitbat} \citep{Batten2019}.

\begin{figure}
    \centering
    \includegraphics[width=\linewidth]{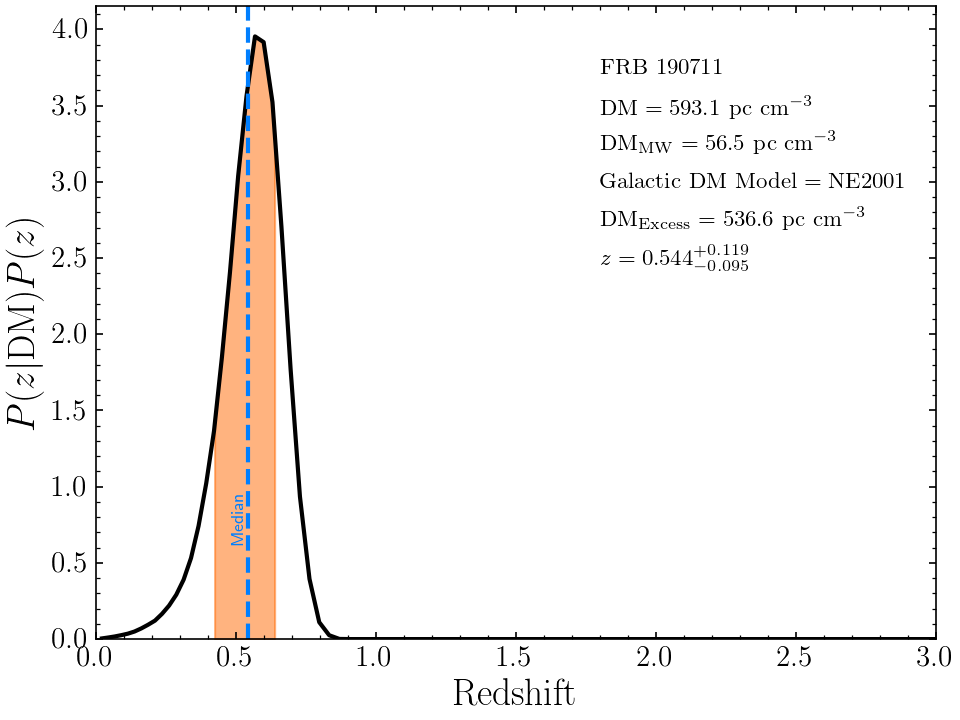}
    \caption{An example redshift PDF created with \textsc{fruitbat} for FRB 190711. The dotted blue line is the median of the redshift PDF and the orange shaded region highlights the 68\% confidence interval.}
    \label{fig:fruitbat_pdf_example}
\end{figure}

In \Cref{fig:fruitbat_pdf_example} we show an example plot of using the $\mathrm{DM}-z$ relation presented in this work to estimate the redshift of FRB~190711. The solid black line is the redshift PDF at $536.6~\mathrm{pc~cm^{-3}}$, the DM excess for FRB~190711. In this case we used a uniform prior ($P(z) = 1$). The filled orange region is the 68\% confidence interval for redshift, and the dotted blue line is the median redshift estimate.

In \Cref{code:fruitbat_example} we show the IPython console input and output that was used to create \Cref{fig:fruitbat_pdf_example} using \textsc{fruitbat}.  

\begin{lstlisting}[language=python, caption={The sample code used to create \Cref{fig:fruitbat_pdf_example}.}, label={code:fruitbat_example},]
In [1]: import fruitbat as frbat

In [2]: frb190711 = frbat.Frb(
          dm=593.1, 
          raj="21:57:40.68", 
          decj="-80:21:28.8", 
          name="FRB\ 190711")

In [3]: frb190711.calc_dm_galaxy(
          model="ne2001")
Out[3]: <Quantity 56.48736954 pc / cm3>

In [4]: frb190711.calc_redshift(
          method="Batten2020")
Out[4]: <Quantity 0.54406244>

In [5]: frb190711.plot_redshift_pdf(
          filename="FRB190711_z_PDF", 
          usetex=True)

\end{lstlisting}

\section{Derivation of DM}
\label{app:dm_derivation}
In this appendix we describe how we calculate DM along lines-of-sight in the \EAGLE\ simulations. 

We begin with the definition of DM. The total DM along a line-of-sight d$l$ to redshift $z$ is as we defined in \Cref{eq:dispersion_measure}
\begin{align}
    \mathrm{DM} = \int_0^z \frac{n_{e, \mathrm{p}}(z)}{1+z}\mathrm{d}l_\mathrm{p} \,,
\end{align}

where $\mathrm{DM}$ is the dispersion measure, $n_{e,\mathrm{p}}$ is the physical electron number density and $\mathrm{d}l_\mathrm{p}$ is the physical distance element. 

Note that here we have added the subscript `$\mathrm{p}$' to $n_{e, \mathrm{p}}$ and $\mathrm{d}l_\mathrm{p}$ to emphasise that these are \emph{physical} quantities. Since the box size of the \EAGLE\ simulations are in comoving units, it is convenient to convert $\mathrm{d}l_\mathrm{p}$ to a comoving quantity (i.e. $\mathrm{d}l_\mathrm{p} = \mathrm{d}l_\mathrm{c} (1+z)^{-1}$). Thus we have  
\begin{equation} \label{eq:DM_derivation_2}
\mathrm{DM} = \int_0^{z)} \frac{n_{e, \mathrm{p}}}{(1 + z')^2} \mathrm{d}l_\mathrm{c}\,.
\end{equation}

In the case of the \EAGLE\ simulations we do not have an infinite sampling of redshift slices, but instead a finite number of simulation boxes with a comoving depth of $L_\mathrm{box, c}$. In practise, we have to break the integral in \Cref{eq:DM_derivation_2} into a sum over the series of boxes
\begin{equation}\label{eq:DM_derivation_3}
\mathrm{DM} = \large\sum_{\mathrm{Boxes}} \frac{n_{e, \mathrm{p}}}{(1 + z)^2}  L_\mathrm{box, c}\,.
\end{equation}

The column density of electrons, $N_e$, along a given line-of-sight is given by $N_e = \int n_{e, \mathrm{p}}(z)\ \mathrm{d}l_\mathrm{p}$. Where $\mathrm{d}l_\mathrm{p}$ is the physical distance element. Therefore the column density through the \EAGLE\ simulation box ($N_{e, \mathrm{EAGLE}}$) with a depth $L_\mathrm{box, c}$ is given by
\begin{equation}
	N_{e, \mathrm{EAGLE}} = n_{e, \mathrm{p}} \frac{L_\mathrm{box, c}}{1+z} \,,
\end{equation}
were $L_\mathrm{box, p} = L_\mathrm{box, c}(1+z)^{-1}$. Through rearranging we find $n_{e, \mathrm{p}} = N_{e, \mathrm{EAGLE}} L_\mathrm{box, c}^{-1} (1 + z)$. 
When we substitute this into \Cref{eq:DM_derivation_3} we obtain:
\begin{equation}
\mathrm{DM} = \large\sum_{\mathrm{Boxes}} \frac{N_{e, \mathrm{EAGLE}}}{(1 + z)}\,. \label{eq:DM_derivation_5}
\end{equation}
Hence, to calculate the DM along lines-of-sight in EAGLE, we can use the column density of electrons, $N_{e, \mathrm{EAGLE}}$, as shown in \Cref{eq:DM_derivation_5}.

\section{Convergence Testing} \label{app:convergence_tests}
In this appendix, we test the convergence of $\left\langle \mathrm{DM_{cosmic}}\right\rangle$, $\sigma_\mathrm{Var}$, $\sigma_\mathrm{CI}$ and $f_\mathrm{NG}$ with box size, simulation resolution and sub-grid physics calibration. The default simulation that we compare to is RefL0100N1504. \Cref{tab:EAGLE_sim} lists the simulations referenced in this appendix.

Convergence testing is necessary to ensure that the results and conclusions measured in this paper are not dependent on artificial properties of the simulation (i.e resolution and box size). One issue that arises is the resolution of subgrid physics models. There are two main approaches: you can use the same subgrid model at high resolution but know the results will be different because the feedback is acting on different scales or you can recalibrate the model for high resolution, but add extra degrees of freedom. In the language of \citet{Schaye2015}, if the simulations we say that the simulation is `strongly converged' if the results do not change with increasing resolution and the subgrid model is held fixed. Alternatively, we say the simulations are `weakly converged' if results do not change with increasing resolution after the subgrid physics model was re-calibrated for its higher resolution. Simulated volume convergence does not suffer from these issues because the mass resolution of the simulation is the same.

We compare the effects of changing box-size by performing the same analysis using two smaller simulations of the same resolution (RefL0025N0376 and RefL0050N0752).

We compare between three 25 cMpc boxes to test the effects of simulation resolution because higher-resolution simulations of 50 or 100 cMpc do not exist. A comparison to RefL0025N0752 and RefL0025N0376 tests the `strong convergence' of the simulations as they are all run with the same subgrid physics parameters. We have also tested the `weak convergence' of by comparing to RecalL0025N0752.  We consider RecalL0025N0752 a better representation of RefL0100N1504 at higher resolution than RefL0025N0752.The strong and weak of simulations is described in section 2.2 of \citet{Schaye2015}.

\subsection{DM - Redshift Relation Convergence}
\begin{figure}
    \centering
    \includegraphics[width=\linewidth]{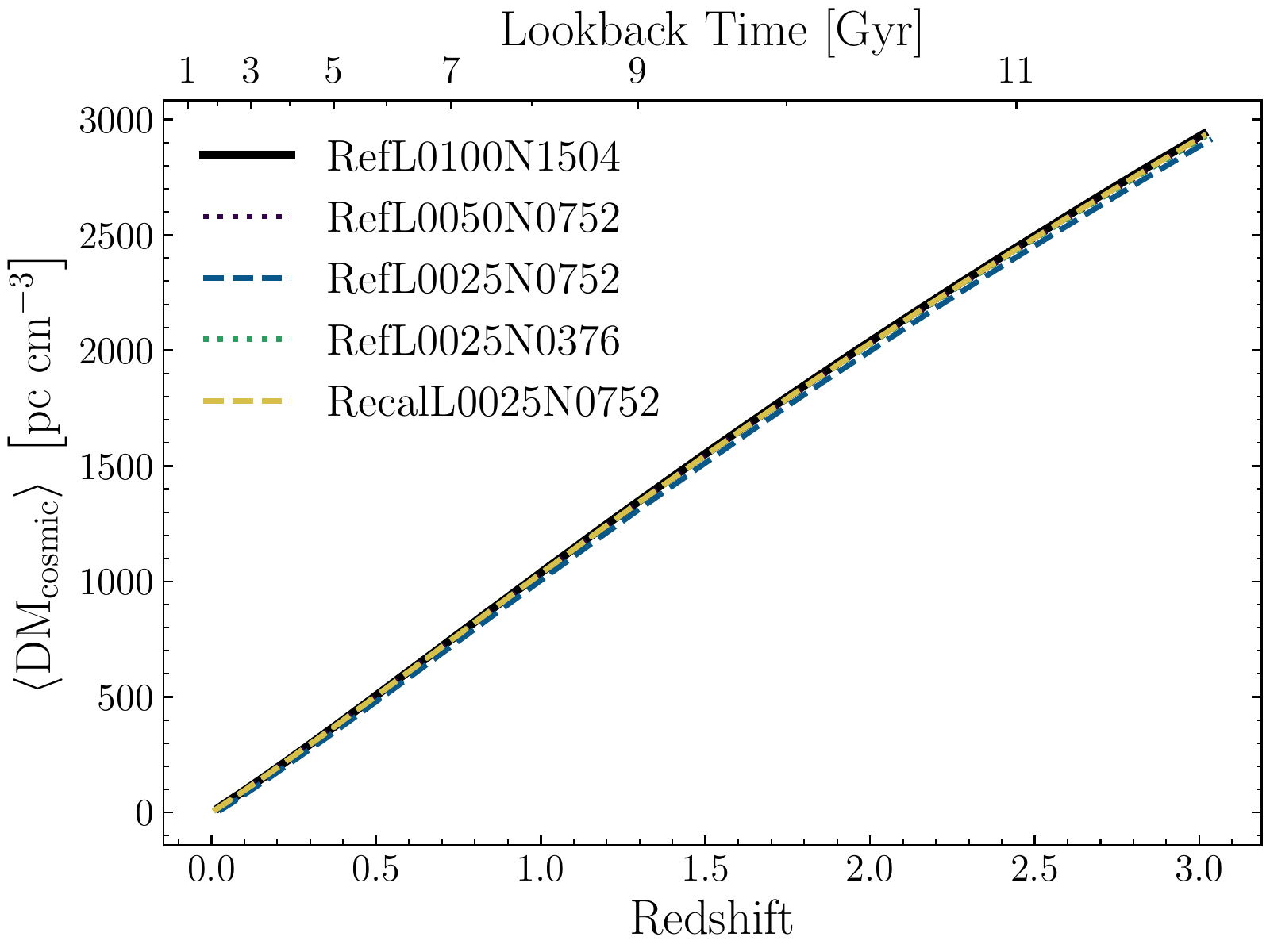}
    \caption{The mean dispersion measure at each redshift calculated using \Cref{eq:DM_Mean} for all the simulations. The simulations with dotted lines (RefL0050N0752 and RefL0025N0376) indicate they have the same resolution as RefL0100N1504 and and differences is due to a box size dependence. The simulations with dashed lines (RefL0025N0752 and RecalL0025N0752) indicate they have the same box size and resolution, but different physics calibrations.}
    \label{fig:conv_test_mean_v_redshift}
\end{figure}

In \Cref{fig:conv_test_mean_v_redshift} we show the effect of box size, resolution and sub-grid physics calibration has on $\left\langle \mathrm{DM_{cosmic}} \right\rangle$. The thick black line is the reference $\left\langle \mathrm{DM_{cosmic}} \right\rangle$ presented in this paper from simulation RefL0100N1504. The dotted lines indicate simulations that have the same resolution, but smaller box size as RefL0100N1504. The dashed lines indicate simulations that have a higher resolution than the reference simulation. We find little to no difference between the $\left\langle \mathrm{DM_{cosmic}} \right\rangle$ for each simulation. This indicates that $\left\langle \mathrm{DM_{cosmic}} \right\rangle$ is extremely well converged in our analysis.  

\subsection{Scatter Convergence}
\begin{figure}
    \centering
    \includegraphics[width=\linewidth]{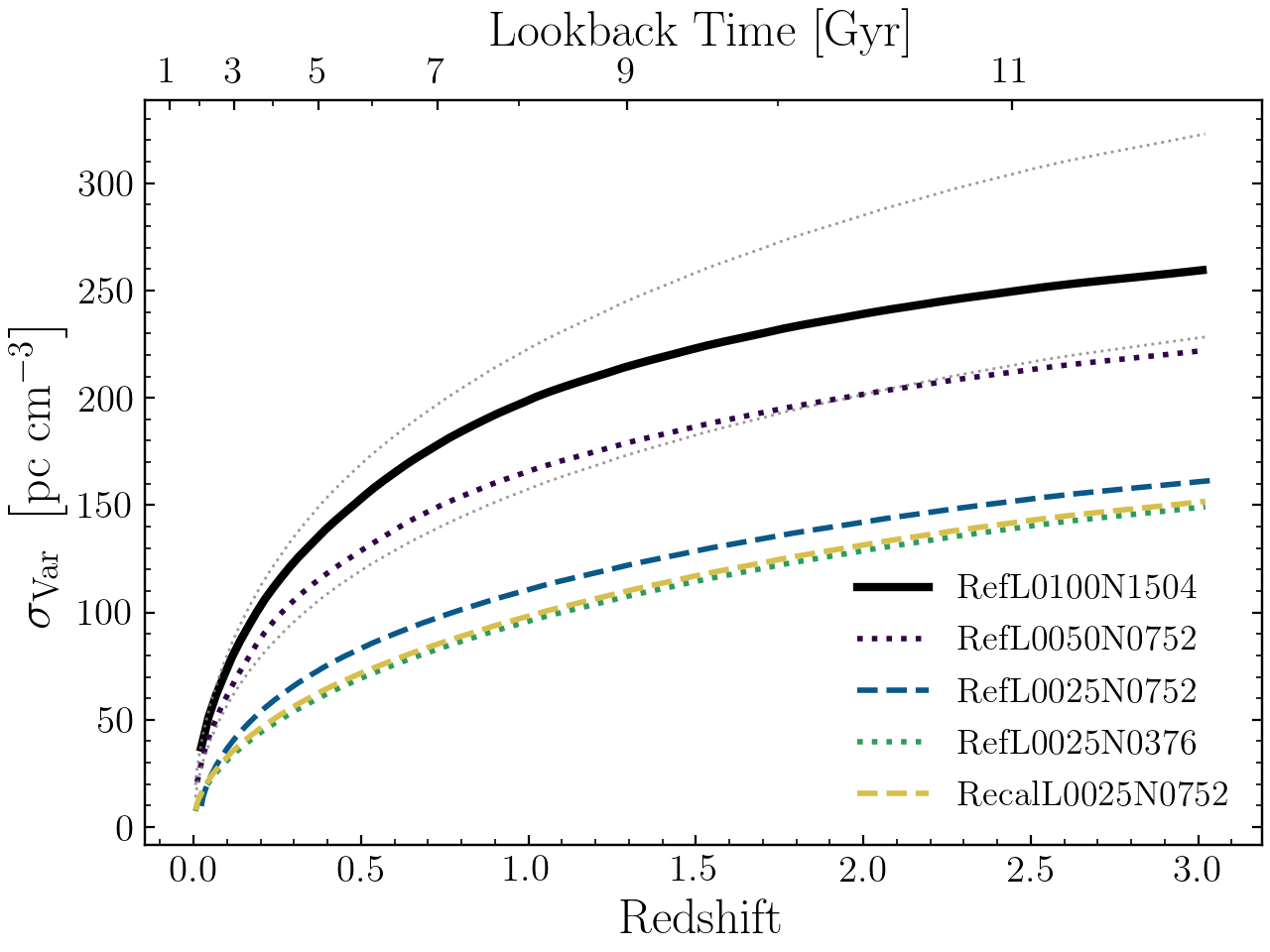}
    \caption{The $\sigma_\mathrm{Var}$ as a function of redshift all the different simulation types. The line styles here are the same as in \Cref{fig:conv_test_mean_v_redshift}. The two light-grey dotted lines are $\sqrt{2}\times \sigma_\mathrm{Var}$ and  $2\times\sigma_\mathrm{Var}$ from RefL0025N0376. For the results to be converged the RefL0100N1504 line should be at or below the highest grey dotted line.}
    \label{fig:conv_test_sigma_var_v_redshift}
\end{figure}

\begin{figure}
    \centering
    \includegraphics[width=\linewidth]{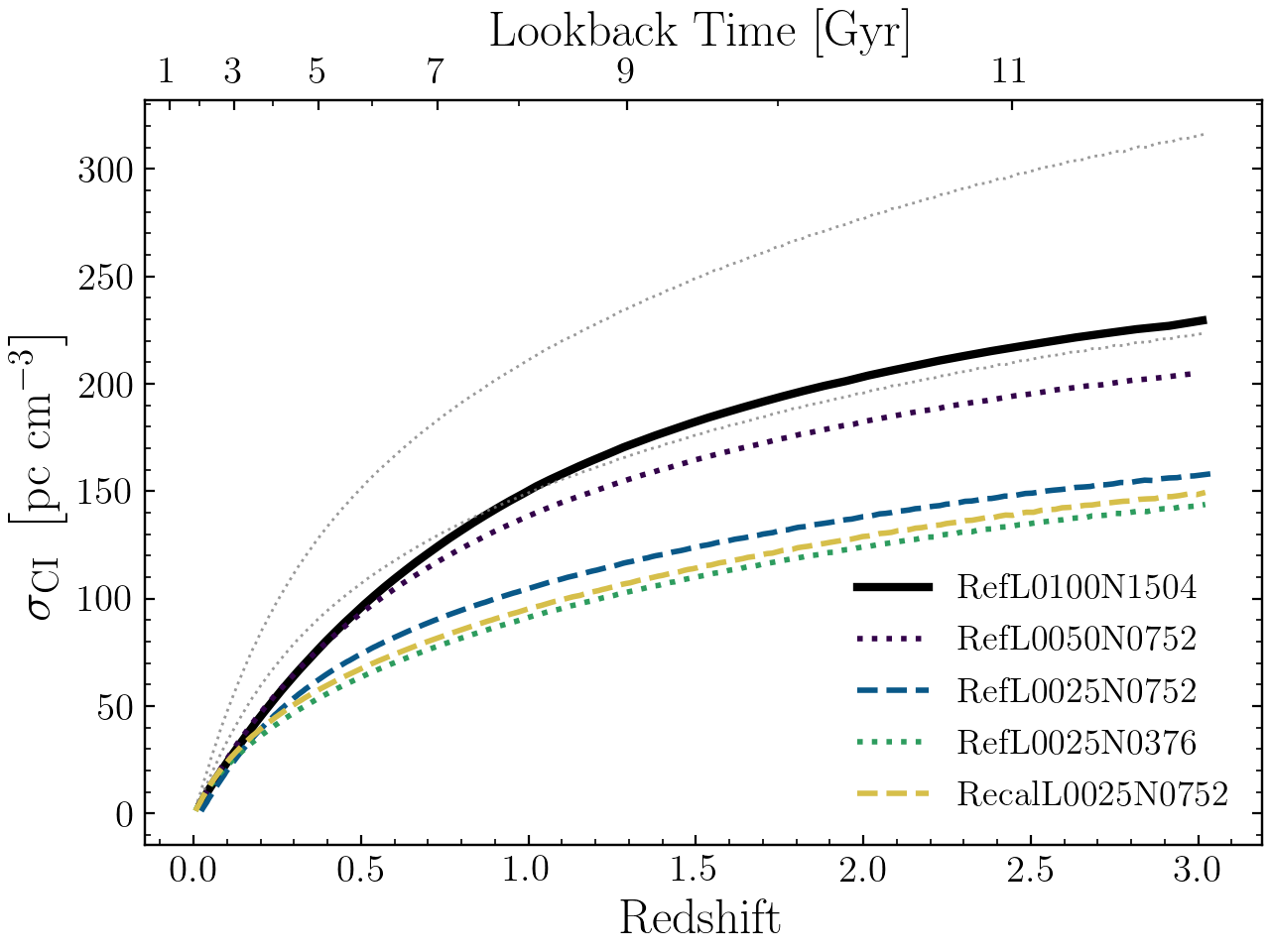}
    \caption{The $\sigma_\mathrm{CI}$ as a function of redshift all the different simulation types. The line styles here are the same as in \Cref{fig:conv_test_mean_v_redshift}. The two light-grey dotted lines are $\sqrt{2}\times \sigma_\mathrm{CI}$ and $2\times\sigma_\mathrm{CI}$ from RefL0025N0376. For the results to be converged the RefL0100N1504 line should be at or below the highest grey dotted line. }
    \label{fig:conv_test_sigma_ci_v_redshift}
\end{figure}

In \Cref{fig:conv_test_sigma_var_v_redshift} and \cref{fig:conv_test_sigma_ci_v_redshift} we plot $\sigma_\mathrm{Var}$ and $\sigma_\mathrm{CI}$ respectively as a function of redshift $z$. The line colours and styles are the same as in \Cref{fig:conv_test_mean_v_redshift}.

We should expect to see some differences between simulations due to the larger number of DM maps used in the smaller simulations.
For example, to satisfy resolution convergence, four 25 cMpc maps from RefL0025N0752 should sum to the same total variance of RefL0100N1504. Since variance is added in quadrature, if we assume that each of the four DM maps of a 25 cMpc simulation are approximately the same variance (which is not necessarily true) then we should expect $\sigma_\mathrm{Var}$ to satisfy 
\begin{equation}
    \sigma_{\mathrm{var},100}^2 \approx 4\sigma_{\mathrm{var},25}^2 \,.
    \label{eq:sigma_var_resolution}
\end{equation}

Here, $\sigma_{\mathrm{var},100}$ and $\sigma_{\mathrm{var},25}$ are the standard deviations of a single 100 cMpc and 25 cMpc DM map respectively. From \Cref{eq:sigma_var_resolution} we would expect to see a $\sqrt{2}$ increase in $\sigma_\mathrm{var}$ for a doubling in simulation box size.

We have over plotted the $\sqrt{2} \times \sigma_\mathrm{var}$ and $2\times\sigma_\mathrm{var}$ as light grey dotted lines in \Cref{fig:conv_test_sigma_var_v_redshift} relative to the RefL0025N0376 line. Similarly we also over plotted $\sqrt{2} \times \sigma_\mathrm{CI}$ and $2\times \sigma_\mathrm{CI}$ in \Cref{fig:conv_test_sigma_ci_v_redshift}. If $\sigma_\mathrm{var}$ and $\sigma_\mathrm{CI}$ are converged with box size, we should expect that the 50 cMpc simulations be below or follow close to the bottom light-grey dotted line and the 100 cMpc simulation to be below or lay close to the upper light-grey dotted line line. We can see here that both the 50~cMpc and the 100~cMpc are less than the appropriate light-grey dotted lines, indicating that we are converged with box size.

Additionally, we see very little difference between the simulations RefL0025N0376, RecalL0025N0752 and RefL0025N0752 indicating we have both strong and weak convergence in resolution.

\subsection{Non-Gaussanity Convergence}
\begin{figure}
    \centering
    \includegraphics[width=\linewidth]{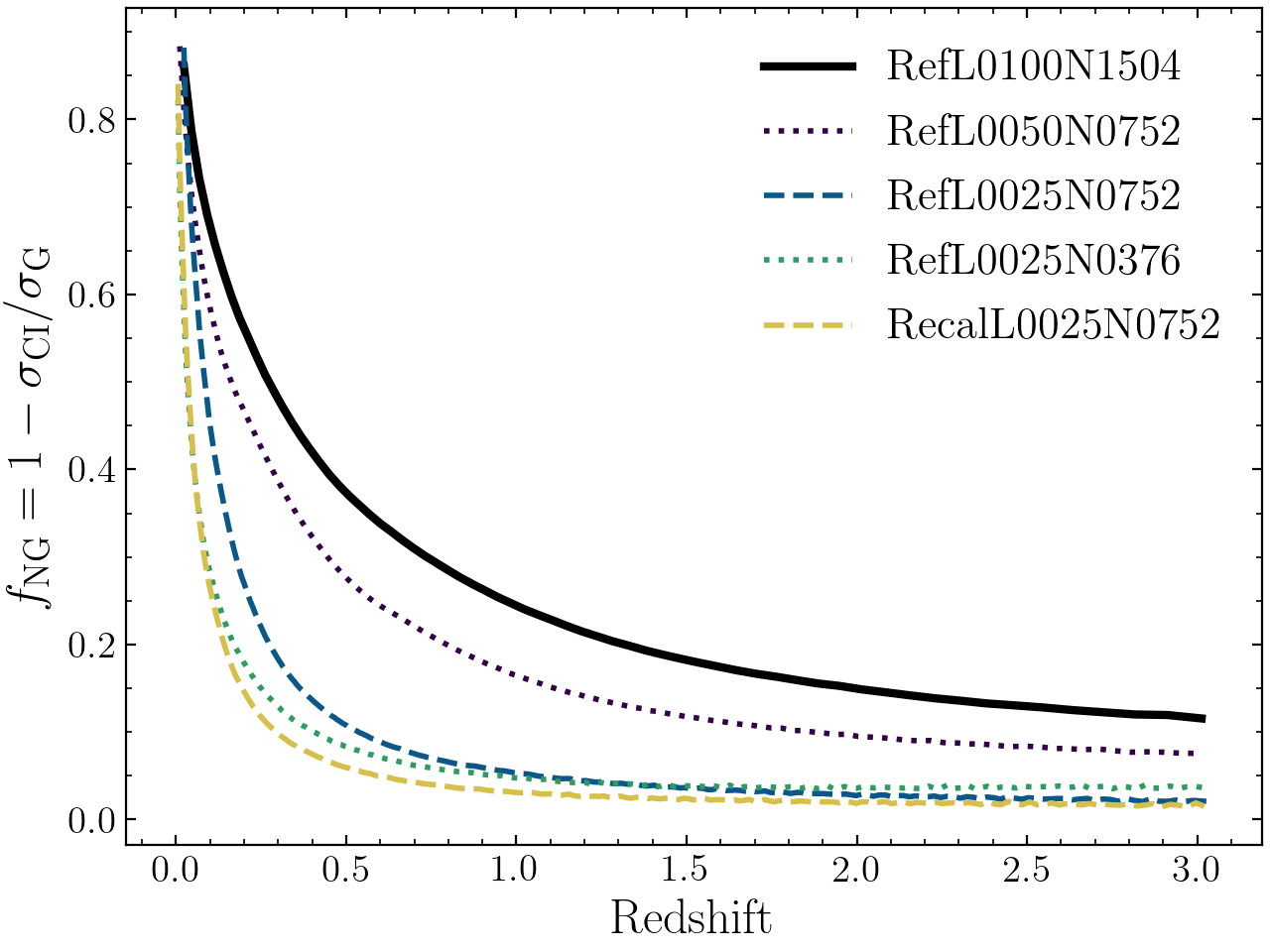}
    \caption{A comparison between $\sigma_\mathrm{Var}$ and $\sigma_\mathrm{CI}$ from \Cref{fig:conv_test_sigma_var_v_redshift,,fig:conv_test_sigma_ci_v_redshift} respectively. A ratio of 0.0 indicates that the $\left\langle \mathrm{DM_{cosmic}}\right\rangle$ PDF at redshift $z$ is Gaussian. Ratios other than 0.0 indicate that the PDF is skewed from normality.}
    \label{fig:conv_test_fng_v_redshift}
\end{figure}

In \Cref{fig:conv_test_fng_v_redshift} we show the effect of box size, resolution and sub-grid physics calibration has on $f_\mathrm{NG}$. The line colours and styles are the same as in \Cref{fig:conv_test_mean_v_redshift}.

The smaller 25 cMpc simulations become close to Gaussian by redshift $z \sim 0.5$. The rate at which the PDFs become Gaussian with redshift in the 25 cMpc simulations is significantly faster than the 50 and 100 cMpc simulations. This is likely because the 25 cMpc simulations are not large enough to contain structure in the extreme tails of the log-normal matter density distribution.

\section{Scramble vs. Transformed Maps Comparison}\label{app:shuffled_vs_transformed}
In this section we compare two techniques that can be used to minimise repeating structure when combining the simulations.

The two techniques we compare are: (i) the technique used in \Cref{sec:shuffle} that we have called the `Scramble Technique' and (ii) a more traditional approach involving rotations, mirrors and translations. We refer to this second approach as the `transformation technique'.

\subsection{Transformation Technique}
For each interpolated DM map except the first (we did not perform any transformations on the redshift $z = 0$ interpolated DM map) we performed a random rotation, mirror and translation. \\

\begin{description}
\item \textit{Rotation}: To rotate the interpolated DM map we randomly chose an angle $\theta$ that is an integer multiple of 90 degrees (i.e 0, 90, 180 or 270 degrees) and rotated the map by $\theta$ in the counter-clockwise direction. We chose $\theta = $ 0, 90, 180 or 270 degrees to ensure that all the pixels in the interpolated DM maps end up aligned across redshifts for ease of computation.
\item \textit{Mirror}: To mirror the interpolated DM map we flip the orientation of the horizontal axis. This has the effect of mirroring the left-right orientation. We chose to only mirror the horizontal axis since mirroring the vertical axis can be achieved through a combination of rotations plus a horizontal mirror. \footnote{A vertical mirror is a 180 degree rotation followed by a horizontal mirror.} It should be noted that we chose the random rotation and mirror in conjunction such that zero-rotation and zero-mirror (relative to snapshot at $z=0$) was not an available option.
\item \textit{Translation}: To translate the interpolated DM map we randomly applied a periodic shift to the rows and columns. We are able to translate the map because \EAGLE\ employs periodic boundary conditions. The number of rows and columns to shift were calculated to ensure that a minimum translation of $10~\mathrm{cMpc}$ occurred. \footnote{There was also a maximum translation of $90~\mathrm{cMpc}$ which corresponds to a $10~\mathrm{cMpc}$ translation in the opposite direction} A translation of $10~\mathrm{cMpc}$ is sufficiently large enough that correlations are nearly absent between large scale structure and hence repetition along line-of-sights should be minimised. 
\end{description}

\begin{figure}
    \centering
    \includegraphics[width=\linewidth]{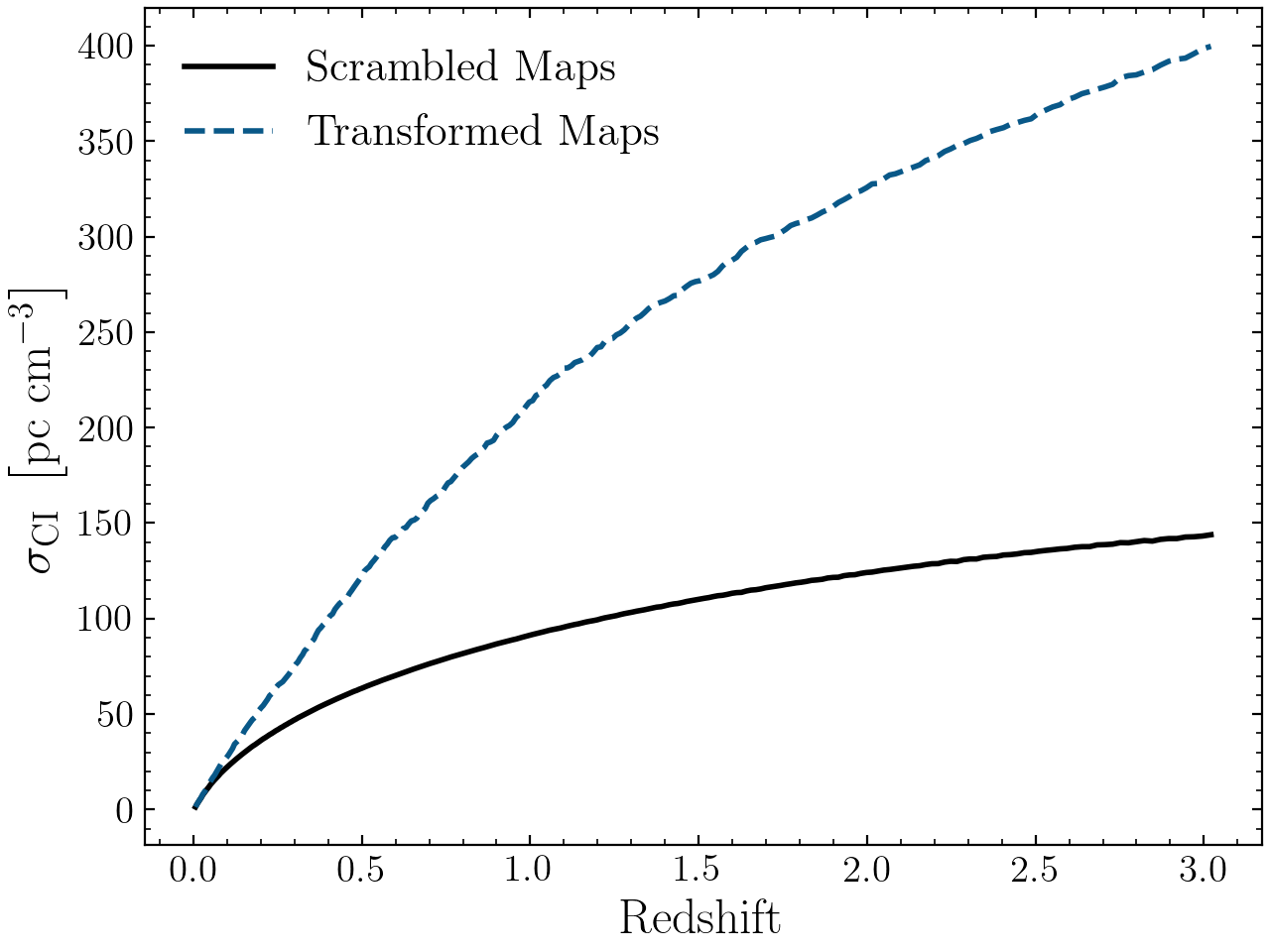} \\
    \includegraphics[width=\linewidth]{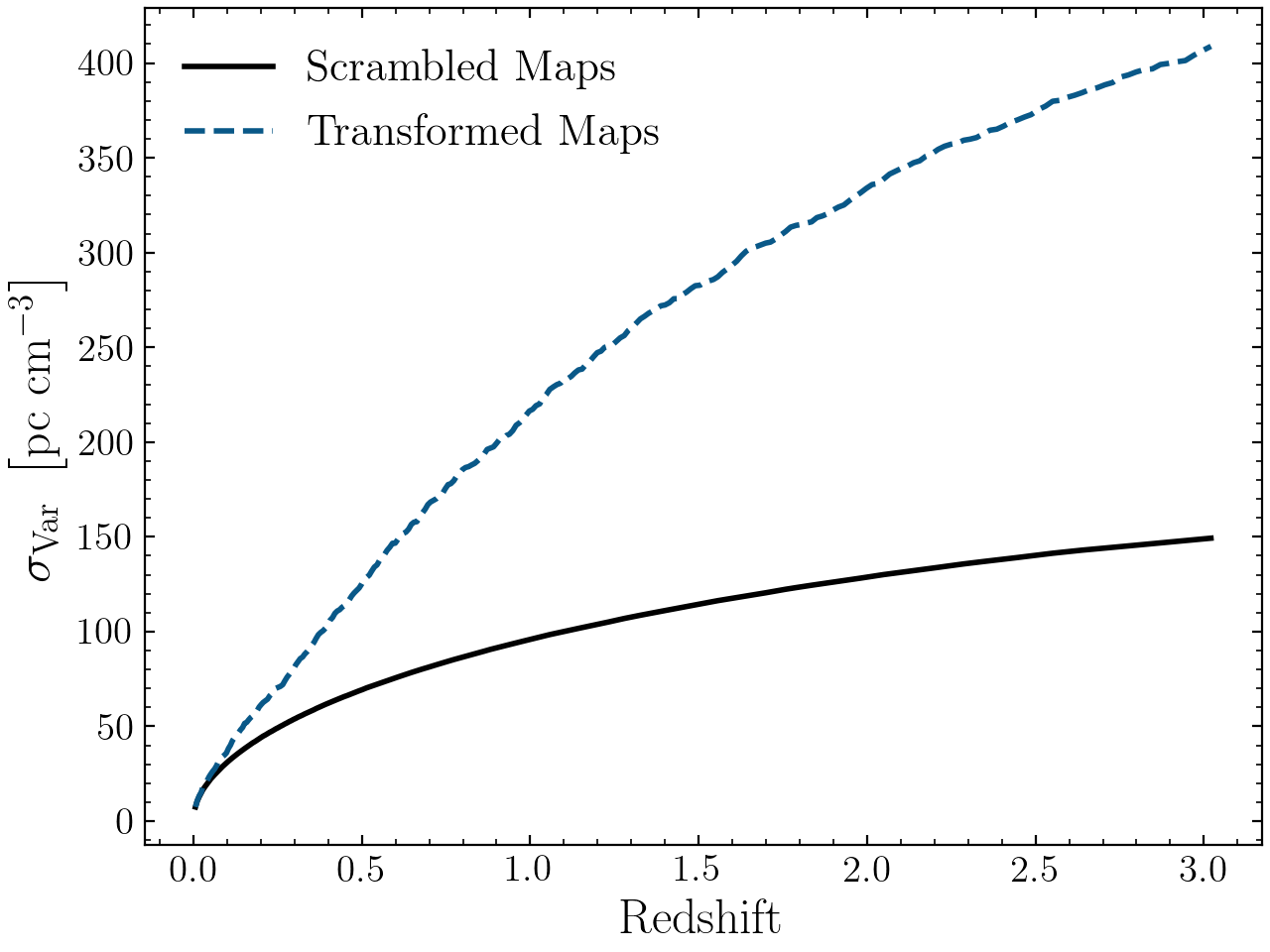}
    \caption{The $\sigma_\mathrm{CI}$ ($\sigma_\mathrm{Var}$) for RefL0025N0376 in the top (bottom) panel comparing the scrambled (solid black) and transformed (dashed blue) techniques. The large difference between the two curves indicates that the transformed technique introduces significant correlations.}
    \label{fig:RefL0025N0376_sigma_scrambled_transformed}
\end{figure}

\begin{figure}
    \centering
    \includegraphics[width=\linewidth]{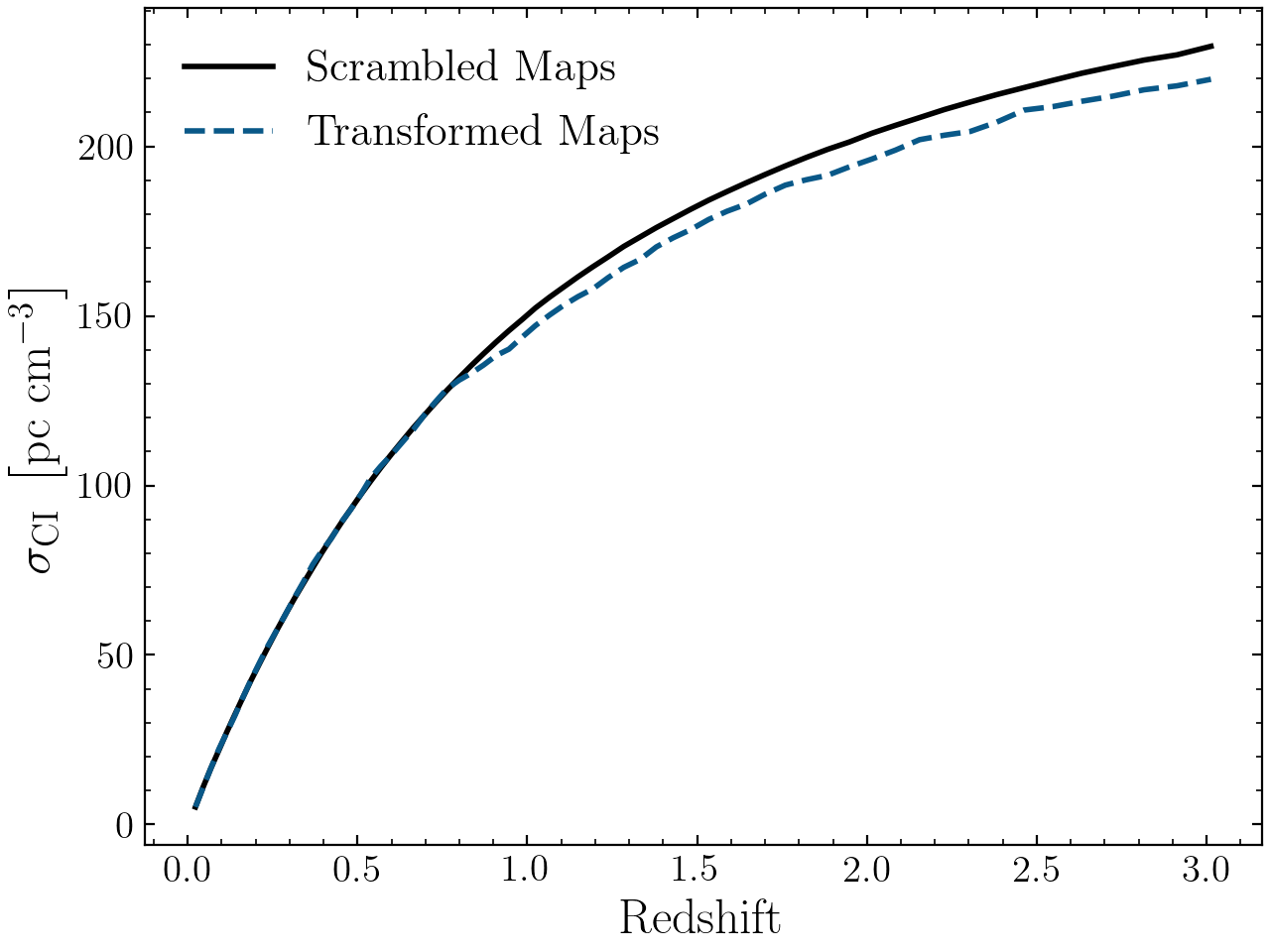} \\
    \includegraphics[width=\linewidth]{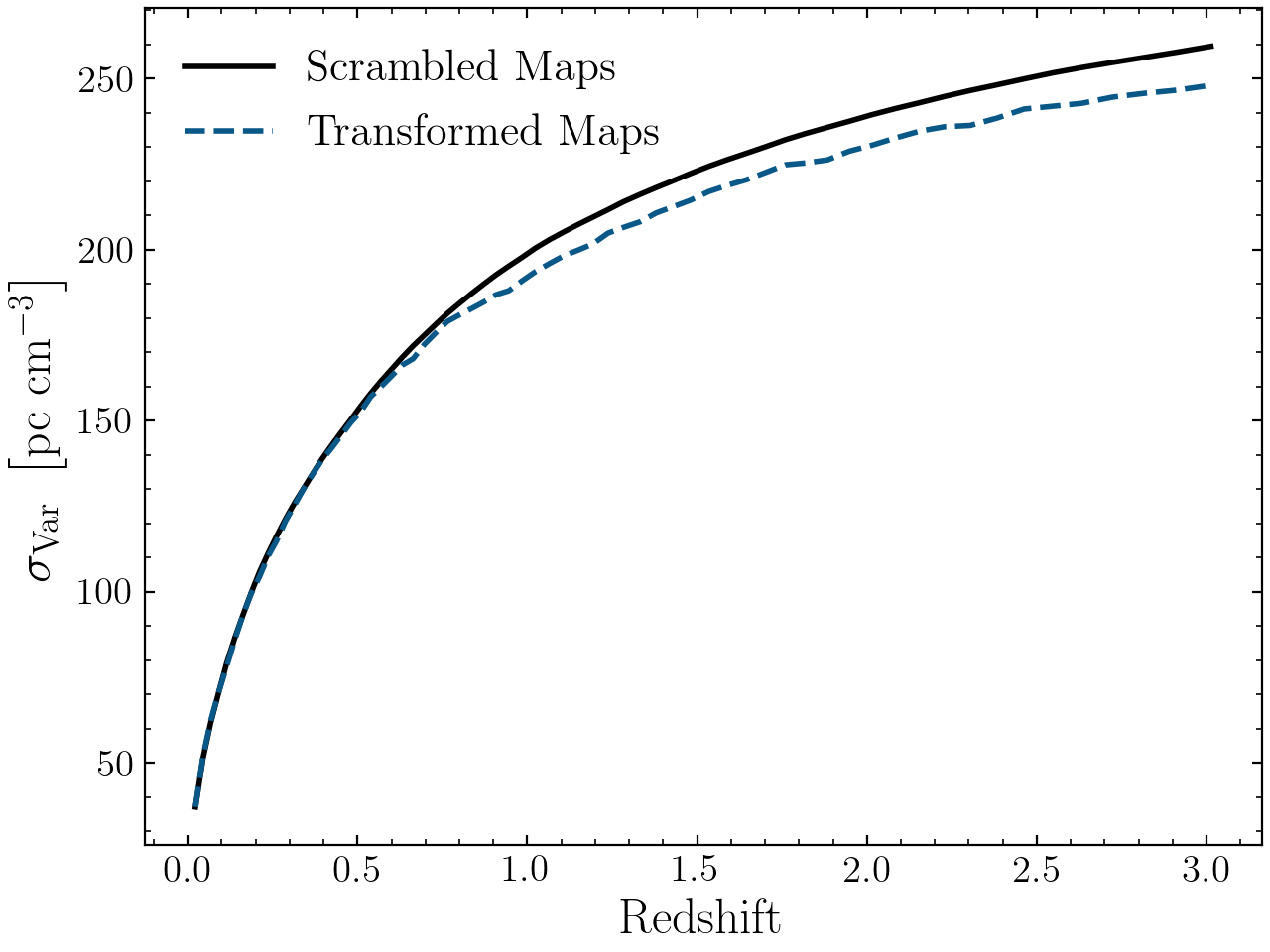}
    \caption{The $\sigma_\mathrm{CI}$ ($\sigma_\mathrm{Var}$) for RefL0100N1504 in the top (bottom) panel comparing the scrambled (solid black) and transformed (dashed blue) techniques. The smaller difference between the two curves as compared to \Cref{fig:RefL0025N0376_sigma_scrambled_transformed} indicates that the transformed technique introduced only small correlations with this size simulation. This suggests that using the transformation technique on boxes that are smaller than 100 cMpc will lead to significant errors on the estimated scatter around the $\mathrm{DM} - z$ relation.}
    \label{fig:RefL0100N1504_sigma_scrambled_transformed}
\end{figure}

We can see that at high redshift the value of  $\sigma_\mathrm{var}$ and $\sigma_\mathrm{CI}$ increases with decreasing box size (i.e. RefL0025N0376, RefL0025N0752 have much larger values than RefL0100N1504). The increased variance is caused by correlations between boxes. 

Since the translation always has a minimum distance of 10 cMpc in both the $X$ and $Y$ directions this means the smaller boxes (i.e. RefL0025N0376, RefL0025N0752, and RecalL0025N0752) do not have a large range of translation space. They only have a small region in the middle to translate to. This coupled with the fact that we require 262 $\times$ 25 cMpc maps to extend out to redshift $z = 3$, causes a significant amount of overlapping structure and correlations.


\bsp	
\label{lastpage}
\end{document}